\documentclass[11pt,a4paper]{elsarticle}
\usepackage[utf8]{inputenc}
\usepackage[english]{babel}
\usepackage{amsmath, bm}
\usepackage{amsfonts}
\usepackage{amssymb}
\usepackage{graphicx}
\usepackage{caption}
\usepackage{subcaption}
\usepackage[left=2cm,right=2cm,top=2cm,bottom=2cm]{geometry}
\usepackage{algorithm}
\usepackage[noend]{algpseudocode}
\usepackage{amssymb}
\usepackage{xcolor}
\usepackage{amsmath}
\usepackage{amssymb}
\usepackage{amsthm}
\usepackage{mathtools}
\usepackage{braket}
\usepackage{bm}
\usepackage{graphicx}
\usepackage{xcolor}
\usepackage{blindtext}
\usepackage{graphicx}
\usepackage{graphics}
\usepackage{verbatim}   
\usepackage{amsfonts}
\usepackage{adjustbox}
\usepackage[toc,page]{appendix}

\usepackage{dsfont}

\DeclareMathOperator{\arcsinh}{arcsinh}

\DeclareRobustCommand*{\square}{{\setlength{\fboxsep}{-.5pt}\fbox{\phantom{l}}}}

\usepackage{siunitx}

\usepackage{enumitem}
\usepackage{lipsum}

\newlength\mylen
\newlist{mycases}{enumerate}{1}
\setlist[mycases,1]{label=\textbf{Case~\arabic*.}, 
  labelwidth=\dimexpr-\mylen-\labelsep\relax,leftmargin=0pt,align=right}

\begin{document}
\begin{frontmatter}

\title{Homogeneous nucleation of dislocations as a  pattern formation phenomenon}

\author[1,2,3]{R. Baggio}
\author[1]{O. U. Salman}
\author[2]{L. Truskinovsky}
\address[1]{LSPM, CNRS UPR3407, Paris Nord Sorbonne Universit\'e, 93400, Villateneuse,  France}
\address[2]{PMMH, CNRS UMR 7636 ESPCI ParisTech, 10 Rue Vauquelin,75005, Paris, France}
\address[3]{UMR SPE 6134, Universit\'e de Corse, CNRS, Campus Grimaldi, 20250, Corte, France}

 \begin{abstract}
Dislocation nucleation in homogeneous crystals initially unfolds  as  a linear symmetry-breaking elastic instability. In the absence of explicit nucleation centers,  such instability develops simultaneously all over the crystal  and   due to the dominance of long range elastic interactions  it  advances   into the nonlinear stage as a collective phenomenon  through pattern formation. In this paper we use a novel mesoscopic tensorial model (MTM) of crystal plasticity to study the delicate role of crystallographic symmetry in the development  of the   dislocation  nucleation patterns  in defect free crystals loaded in a hard device. The model is formulated in 2D and we systematically compare lattices with square and triangular symmetry. To avoid the prevalence of the conventional   plastic mechanisms, we consider the   loading paths represented by pure shears applied on the boundary of the otherwise unloaded body. These loading protocols can be qualified as exploiting the 'softest' and the 'hardest' directions and we show that the associated dislocation patterns are strikingly different.
\end{abstract}

\begin{keyword}
crystal plasticity \sep dislocation nucleation \sep lattice invariant shear \sep homogenenous nucleation \sep pattern formation \sep mechanical twinning

\end{keyword}

\end{frontmatter}


\section{Introduction}
\label{sec:Intro}

Plastic flow in crystals is a result of the  motion of crystal   defects among which  the dominant role is   played by  lattice dislocations  \cite{Movchan1998-jc,Karlin2000-fb,Movchan1987-sz,Bullough2020-hr,Movchan2003-sn}. Understanding the mechanism of creation  of dislocations is essential for the development of the fundamental theory of  crystal plasticity allowing one to control the  mechanical strength of crystalline materials \cite{Geslin2017-fb,Mayer2021-bw,Lilleodden2003-tu,Mason2006-sd}.

Homogeneous nucleation of dislocations in crystalline solids   attracts particular   attention as the main  mechanism for incipient plasticity in nanomaterials  where one usually has to deal with practically defect-free crystals \cite{Aubry2011-vx,Asenjo2006-lk,Zhang2020-ax,Skogvoll2021-jt}. Since the action of standard (heterogeneous) dislocation sources   at these scales   is suppressed,  the  knowledge of alternative (homogeneous)  dislocation nucleation mechanisms is of crucial importance for the understanding   of the response of such  materials   which are known to demonstrate   extraordinary  mechanical properties due to the presence of  peculiar, micro-scale-specific deformation mechanisms  
 \cite{Li2007-pp,Zhu2009-yx,Li2010-pr}.

Nucleation of dislocations signals  the  loss of stability of  a perfect lattice subjected to sufficiently large shear stresses \cite{Grimvall2012-lh}. The  resulting symmetry breaking  instability may lead to  reconfiguration of only  few   atomic bonds, as is the case of a   nucleation of a single dislocation, or carry a large-scale restructuring of the  atomic lattice, as during  a catastrophic, brittle-like, collective nucleation of a large number of dislocations which leads to the formation of intricate dislocation patterns \cite{Salman2021-sn}. Although at macro-scales such massive  nucleation of dislocations  can be  usually neglected  in comparison with    emission of individual dislocations from heterogeneities, it may be also  a  dominant factor in bulk materials subjected to high intensity dynamic loadings \cite{Shehadeh2016-mt,Bringa2006-yu}.

Given that   the  sizes associated with dislocation cores can be as small as a few lattice spacings, the  continuum theory is  hardly applicable for the description of  the developed (post-bifurcational) stages of   lattice  instability resulting in the formation of  dislocations.  Therefore molecular dynamics simulation  played an important role in uncovering the fundamental mechanisms of the nucleation of individual dislocations, however  its limited timescale still remains a significant challenge for studying collective nucleation at experimentally relevant conditions \cite{Bulatov2006-uz}. Therefore  various accelerated meso-scale  approaches  have been used including the  microscopic phase-field crystal theory  \cite{Salvalaglio2020-eb,Chan2010-qt},
  the  multi-scale quasi-continuum method \cite{Shenoy1999-gh}, the periodized-discrete-elasticity model \cite{Plans2007-cx},  and   the phase-field dislocation dynamics  \cite{Javanbakht2016-dr}. Each of these  conceptual and computational approaches was successful in addressing  a particular range of time and length scales.  
    

Major efforts have been  focused on finding the dislocation nucleation   criterion  \cite{Li2002-uj,Van_Vliet2003-yg,Miller2004-bm}. Given that    behind dislocation nucleation is  a linear  instability of an elastically pre-stressed  solid,  
many attempts were made  to reduce  the corresponding  continuum-scale criterion  to   nanoscale, for instance, by using the continuum loss of strong ellipticity condition    with  atomic level entries    \cite{Garg2015-ld,Delph2009-xj}.   However, even in the case of apparently homogeneous dislocation nucleation under micro-indenter, the molecular dynamics simulations    revealed  complex mesoscale   processes involving    a large number of  atoms and producing   a strong local distortion  of the lattice which  makes   a phonon stability analysis hardly applicable \cite{Schall2006-jc,Tschopp2007-dm,Miller2008-jk,Wagner2008-ed}.  As a result various nonlocal corrections were  proposed to 'delocalize' the  mesoscale atomic acoustic tensor  and the results  were  extensively compared with  molecular dynamics  simulations  \cite{Garg2016-kz}.  Despite this progress, our ability to predict the instant and the location of the nucleation of an individual dislocation  remains  limited,  while  the first efforts to understand  the corresponding collective effects  have started only recently \cite{Baggio2019-rs,Salman2021-sn}.  Moreover, little remains known about the collective side of   dislocation nucleation including the dependence of emerging patterns of cells and walls on the crystallographic symmetry of the lattice. 
    
The goal of  this paper is to contribute to the understanding of the  collective nucleation of dislocations in perfect crystals as a bifurcation phenomenon with the focus on post-bifurcational development of patterns and textures.   We assume that in the absence of explicit nucleation centers,  the implied  instability develops simultaneously all over the crystal and   that, due to the dominance of long range elastic interactions,  it  proceeds  into the nonlinear stage as a cooperative avalanche which involves self-organization of dislocations into energy minimizing patterns. We design a series of numerical experiments  where we load   pristine  crystals with different crystallographic symmetries beyond the stability limit of the homogeneous state and then  study the  transient unfolding of the dislocation nucleation avalanche which leads to the catastrophic stress drop as the optimal dislocational microstructure settles down.  For simplicity we operate in  2D  where  we can systematically compare the peculiarities of the collective nucleation   in lattices with square and triangular symmetry. To avoid immediate activation of the conventional   plastic mechanisms, we consider the   loading paths represented by  pure  shears applied on the boundary of the otherwise unloaded body. These loading protocols can be qualified as exploiting the 'softest' and the 'hardest' directions and we show that the associated dislocation patterns are strikingly different.

Our main computational tool is  the  novel mesoscopic tensorial model (MTM) of crystal plasticity allowing one   to capture in a geometrically precise way   the   role  of  crystallographically-specific lattice invariant shears  while still operating with the macroscopic notions of stress and strain \cite{Salman2011-jd, Salman2012-us, Baggio2019-rs, Salman2019-no,Salman2021-sn,baggio-arxiv}. The model implies  the construction of an energy density respecting the global symmetry of Bravais lattices  described by the group $GL(n,\mathbb{Z})$ \cite{Ericksen1970-rx,Ericksen_undated-tm,Folkins1991-bv,Parry1998-sv,Conti2004-sv}.

The resulting theory can be viewed as a finite element version of nonlinear elasticity theory accounting for geometrically nonlinear kinematics. The size of the elements is viewed as a physical regularizing (cut-off) parameter bringing an internal scale into the theory. Behind such coarse-grained approach lies the assumption that the deformation inside the meso-scale material elements can be considered as affine and their response is characterized by an  effective energy landscape which is globally periodic due to the presence of an infinite number of equivalent lattice configurations. From the perspective of such Landau-type continuum theory, plastically deformed crystal can be seen as a multi-phase mixture of equivalent “phases”. Plastic yield can be then interpreted as an escape from the reference energy well, and plastic “mechanisms” can be linked to low-barrier valleys of the energy landscape. Rate-independent dissipation emerges in such theory due to the fast (abrupt, at the time scale of the loading) well-switching events describing elementary plastic slips.  
 
The main advantage of the MTM approach is that it is formulated in terms of macroscopically measurable quantities (stress and strain) while being able to distinguish between different crystal symmetries including the resolution of the symmetry dependent  configuration  of the dislocation cores. It can therefore account adequately for both long- and short-range interactions between dislocations. Most importantly, it allows for topological transitions associated with dislocation nucleation and annihilation even though the details of the corresponding “reactions” may appear as blurred on the scale of regularization. Last but not least, in the MTM approach the interaction of dislocations with various obstacles, including self locking and the formation of other types of dislocational  entanglements  can be handled without introducing ad-hoc relations.
  
Using  this modeling framework we show  that   following the loss of elastic stability plasticity develops in the form of a system size avalanche which involves  massive nucleation of dislocations which self-organize into system size patterns. The latter   involves the formation of extended low-energy patches (or grains) undergoing pseudo-rigid rotations. Individual grains   are separated by high-energy dislocation walls.  
The observed   deformation patterns  defy conventional continuum description with its insistence on rigid plastic mechanisms limited to crystallographically specific simple shears and the neglect of the effects of geometrical nonlinearity. More complex picture is observed with various slip systems activated simultaneously and finite elasticity playing an important role in the  observed dislocation patterning. 

 
 The fact that the MTM energy can be formulated for lattices with different symmetries and that we can model  general loading paths  allows us to explore non-trivial deformation mechanisms peculiar to lattices with higher and lower  symmetries.   To highlight these ideas we  focus  in what follows on  the simplest nontrivial case of 2D lattices with two types of symmetries, square and triangular.  We study systematically  two fundamentally  different loading directions  which we consider as providing conceptual bounds for the  whole spectrum of available   responses. One of them is directed towards the   lowest and another one to the highest energy barrier   away from the original energy well. The resulting breakdown of the original homogeneous state  displays complex nucleation pattern with a large number of nucleated dislocations forming a highly organized crystal texture.  The 'softest'  path  highlights the role of the metastable  phases in driving the complexity of the emerging dislocation arrangement. The 'hardest'  path shows in some cases the possibility of collective rearrangements of the lattice taking the form of inelastic rotations  in which dislocations play  the role of invisible intermediaries. 
 

The paper is organized as follows. We begin  by introducing the $GL(2,\mathbb{Z})$-invariant energy and discuss the resulting energy landscape (Section \ref{sec:mesoen}). In Section \ref{sec:staban}, we propose  the criterion detecting  the   instability of the homogeneously loaded lattice  which reveals various  features of the activated instability modes. We then present in Section \ref{sec:simres} the results of the numerical experiments  which confirm the validity of our instability criterion and show  the  post avalanche arrangement of the nucleated dislocations.  A brief description of the numerical method  is given in the Appendix. Our conclusions are summarized in the final Section \ref{sec:conc}.

\section{The  model}
\label{sec:mesoen}

\paragraph{Lattice invariant shears.}The proposed model, whose simplest nontrivial formulation is   for  2D Bravais lattices which are solely  considered in this paper,   allows one  to include plastic deformation in a continuum elastic framework, while simultaneously accounting of the discrete nature of the underlying lattice structure. This is achieved with the construction of an energy density whose material symmetry properties are described by the global symmetry group of the lattice $GL(2,\mathbb{Z})$. 
The latter is broader than  the crystallographic point group \cite{truesdell2004non} and includes 
  the  lattice invariant shears  accounting for plastic slips  \cite{ericksen1977special,ericksen1979symmetry,ericksen1980some,ericksen1987twinning,ericksen1991weak,ericksen2005cauchy}.


The   energy density in the MTM model should be invariant of the action of the group $GL(2,\mathbb{Z})$ which is comprised  of unimodular integer valued matrices $\mathbf{m}$. Indeed, two basis  ${\bf e}_I$ and  ${\bar{\bf e}}_I$ describe the same lattice if and only if \cite{Pitteri2002-rm}:  
$
{\bf {e}}_J  =  m_{IJ}  \bar{\bf e}_I \;\; \mbox{with} \;\; { m}_{IJ} \in {\mathbb Z}.
$
Then, we can say that all 2D simple lattices are invariant under the action of a group
$
GL(2,\mathbb{Z}) = \left\{{\bf m},\, m_{IJ}\in{\mathbb Z},\, \det(\bf m)=\pm1 \right\}.
$
The fact that this matrices are unimodular (i.e. $\det{\bf m}=\pm1$) reflects the condition that these transformations do no affect the volume of the lattice cell (the case $\det{\bf m}=-1$ corresponds to reflection). 
We remark that the  group $GL(2,\mathbb{Z})$ accounts for 
the lattice invariance in shear, but also of invariance under 
rotations and reflections  and in this sense the group $GL(2,\mathbb{Z})$  constitutes the  finite strain extension  of the  crystallographic point group \cite{pitteri1984reconciliation}. 
 Every time we multiply a lattice basis with a matrix $\mathbf{m} \in GL(2,\mathbb{Z})$, we obtain   a crystallographically-equivalent structure with exactly the same energy. The resulting multiplicity of the energy wells implies that  such equivalent configurations can be  interpreted as different "phases" describing the same crystal. In such a description, dislocations will appear  as  incompatible parts of the resulting  'phase  boundaries'. 
 
 In the following we take  for granted that the lattice energy density $\varphi\left( \bar{\mathbf{e}}_i\right)$, where $ \bar{\mathbf{e}}_i=\mathbf{F}\mathbf{e}^0_i$ is the  deformed basis  while   $\mathbf{e}^0_i$ is the reference  basis, can be identified with a continuum strain energy density such that $\phi(\mathbf{F}):=\varphi(\mathbf{F}\mathbf{e}^0)$, with $\mathbf{F}=\nabla\mathbf{u}$ the deformation gradient. 
In view of frame indifference requirement,  the strain energy density $\phi$ must  be a function of the lattice metric tensor ${\bf C}={\bf F}^T{\bf F}$~\cite{Finel2010-zw,Salman2019-cg}. The configuration space is then  described by the three significant components of the metric tensor: $C_{11}, C_{22}$ and $C_{12}$. 
Every point of the surface $\det{\bf C}=C_{22}C_{11}-C^2_{12}=1$ corresponds to an orbit represented by rigidly  rotated lattice configurations. 
\begin{figure*}
\centering
\includegraphics[width=.8\textwidth]{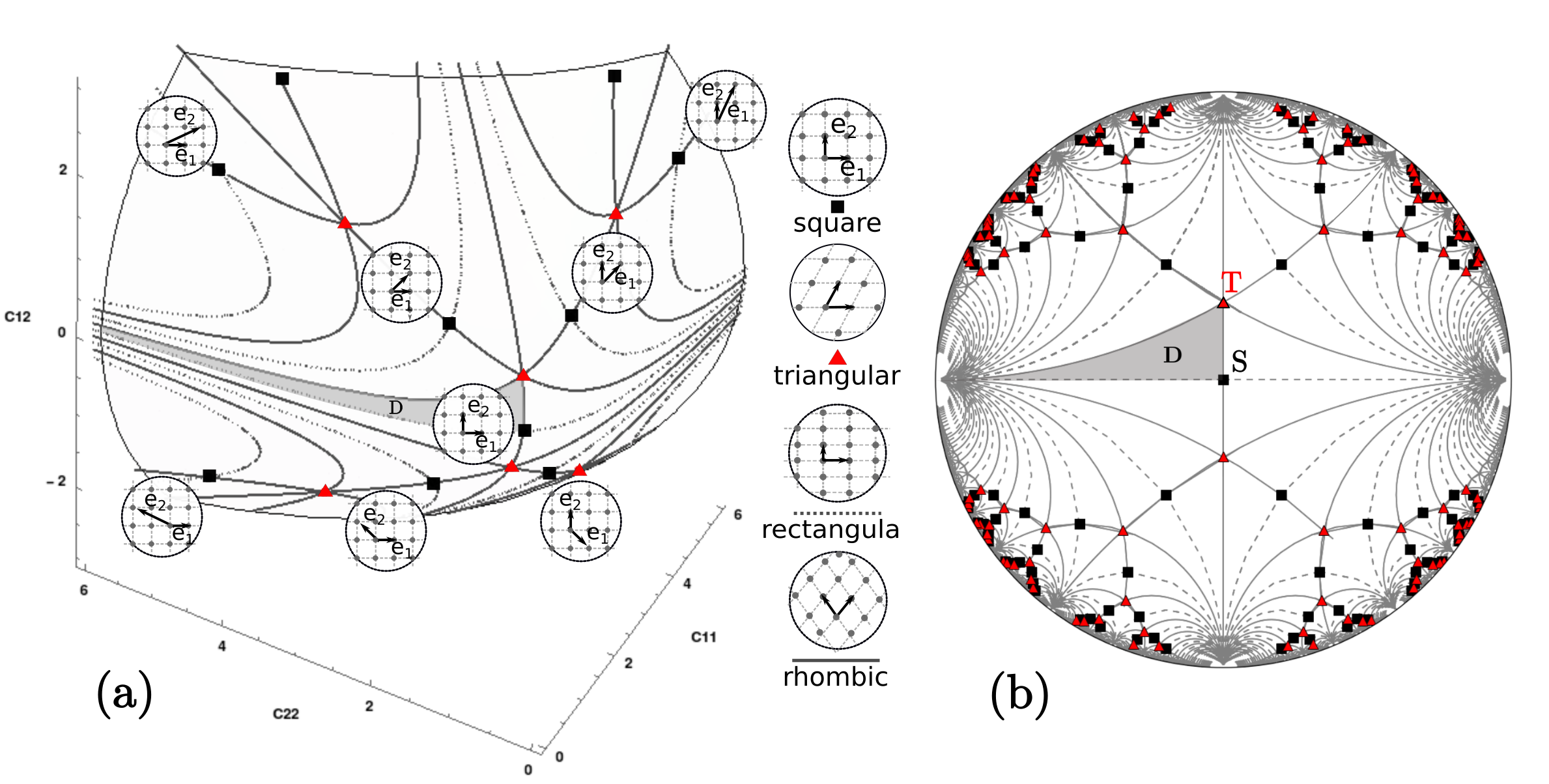}
\caption{\footnotesize{ {\bf (a)} A portion of the hyperbolic surface $\det{\bf C}=1$, points and lines corresponds to the different Bravais lattices. The respective basis vectors ${\bf e}_i$ are shown on insets. {\bf (b)} A stereographic projection  of the   surface $\det{\bf C}=1$  on the Poincare disk. $D$ is the minimal periodicity domain. }} \label{fig:atlas}
\end{figure*}

\paragraph{Minimum periodic domain.} The global  invariance  of the energy suggests that we can  construct  the image of ${\bf C}$ in the minimum periodicity domain
$
D =  \{C \in \det{\bf C}=1, \quad
   0<C_{11}\le C_{22},\quad 0\le C_{12}\le C_{11}/2\}.
$
The metric tensors  belonging to it are associated with lattices basis characterized by  the  "minimal"  vectors $\tilde{\mathbf{e}}_1,\tilde{\mathbf{e}}_2$, because they are  selected in such a way   that:
 $\tilde{\mathbf{e}}_1$ is the shortest lattice vector and 
 $\tilde{\mathbf{e}}_2$ is the shortest lattice vector not collinear with $\tilde{\mathbf{e}}_1$ and for which  the sign is chosen in such a way that the angle between the two is acute.  This type of basis is said to have reduced form of Lagrange  \cite{engel2012geometric}. 
 
To better visualize the tessellation of the configurational space into equivalent periodicity domains, we will use in what follows the stereographic projection of  the infinite   surface $\det{\bf C}=1$ on a disk with unit radius (Poincar\'e disk). The mapping, which  associates the configuration $(C_{11},C_{22},C_{12})$ with the  point $(x,y)$ on the unit disk, is given by the formulas  
\begin{eqnarray}
x=\frac{(\frac{C_{12}}{C_{22}})^{2}+(\frac{\sqrt{\det \bf C}}{C_{22}})^{2}-1}{(\frac{C_{12}}{C_{22}})^{2}+((\frac{\sqrt{\det \bf C}}{C_{22}})+1)^{2}},\,\,                    
y=\frac{2(\frac{C_{12}}{C_{22}})}{(\frac{C_{12}}{C_{22}})^{2}+((\frac{\sqrt{\det \bf C}}{C_{22}})+1)^{2}}.
\end{eqnarray} 
In Fig. \ref{fig:atlas}   we show the location of the minimal periodicity domain $D$ on the hyperbolic surface $\det{\bf C}= 1$ in the space of metric tensors and on its projection on the Poincar\'e disk. We highlight there the  configurations  $\bf S$ and $\bf T$ which  are the unique representatives of the infinite equivalence classes of unloaded square and triangular lattices, belonging to  $D$. The  small black squares in Fig. \ref{fig:atlas}   corresponding to other (not belonging to $D$) variants of the square lattice,  while  the other  equivalent  variants of triangular lattices with hexagonal symmetry  are represented by small red triangles. The rectangular and the rhombic lattices  with one parametric degeneracy are also located in Fig. \ref{fig:atlas} along the  continuous and dashed grey lines;  the generic  obliques lattices with two parametric degeneracy  are located in the open  regions of the configuration.

\paragraph{Lagrange reduction.} For the 'equivalent'  of ${\bf C}$ inside the minimal periodicity domain we use the notation   ${\bf C^0}$.  The metric tensor  ${\bf C^0}$ is defined by the mapping  
$
       {\bf C}^0 = {\bf m}^T{\bf C} {\bf m}$.
%
The task of finding the corresponding unimodular matrix   ${\bf m} $ is known as the   Lagrange reduction \cite{engel2012geometric}. 
  It is a recursive procedure  which can be formulated in the form of an algorithm \cite{engel2012geometric}:
%
(i) initiate ${\bf m} =  \mathbb{I}$; 
(ii) define the following three matrices : ${\bf m}_1=\begin{pmatrix}
1 & 0 \\
0 & -1 
\end{pmatrix}$, ${\bf m}_2=\begin{pmatrix}
0 & 1 \\
1 & 0 
\end{pmatrix}$, ${\bf m}_3=\begin{pmatrix}
1 & -1 \\
0 & 1 
\end{pmatrix}$; 
(iii) initiate recursive algorithm : 
(iv) if $C_{12}<0$,   change sign to $C_{12}$, ${\bf m}\rightarrow ${\bf m}${\bf m}_1$; 
(v) if $C_{22}<C_{11}$, swap these two components, ${\bf m}\rightarrow ${\bf m}${\bf m}_2$;
(vi) if $2C_{12}>C_{11}$,  set $C_{12}=C_{12}-C_{11}$, and  $C_{22}=C_{22}+C_{11}-2C_{12}$, ${\bf m}\rightarrow ${\bf m}${\bf m}_3$.  
Note  that the action of the matrix  ${\bf m}_1$ is related to the sign of the angle between two lattice vectors $\mathbf{e}_i$ and returns an acute angle, whereas the action of the matrix  ${\bf m}_2$ is to swap two  lattice vectors $\mathbf{e}_i$. Therefore, both these two operations do not result in any change in vectors' length and effectively  propagate the metric in the same elastic well composed of the four copies of the fundamental domain $D$ and therefore are not associated with a plastic strain. On the other hand, the length of the lattice vectors is changed (shortened) under the action of the matrix  ${\bf m}_3$, which indicates that the current metric belongs to another elastic well and accumulates plastic strain.

\paragraph{Energy density.} Given that  the energy density will  be  defined fully as long as it is   defined  in the minimum periodicity domain 
and we will use for such a single period description a special notation   
$\phi_D( \bf C^0)$ so that 
 $ \phi({\bf C})= \phi({\bf m}^T{\bf C} {\bf m}) = \phi_D( \bf C^0).$
By defining $\phi_D$ as a function of scaled variables $ \tilde {\bf C}={\bf C}/({\det^{1/2}\bf C})$   we   decouple the  isochoric  contribution to the energy from the  volumetric one that can be added separately. We will require $\phi_D$ to satisfy $\mathcal{C}^2 $ smoothness, which ensures the continuity of the elastic moduli. Moreover, $\phi_D$ must have a minimum which corresponds to the chosen crystal symmetry. For instance, when modelling a square lattice,  $\phi_D$ will be constructed in such a way that minimum coincides with the square symmetry lattice (that is point $C_{11}=C_{22}=1$, $C_{12}=0$). 

A general $6$-th order polynomial energy $\phi_D$ with the required properties was introduced in  \cite{Conti2004-sv}. 
%
The energy density is written in terms  of the three invariants:
$I_1 = \frac{1}{3} (C_{11} + C_{22} - C_{12})$, $ 
I_2= \frac{1}{4} (C_{11} - C_{22})^2 + \frac{1}{12}(C_{11} + C_{22} -
4 C_{12})^2$ and $                   
I_3= (C_{11} - C_{22})^2 (C_{11} + C_{22} - 4 C_{12}) - \frac{1}{9} (C_{11} + C_{22} -
4 C_{12})^3$
and can be written as 
$
  \tilde\phi_D(\tilde {\bf C}) =  \beta_1 \psi_1 (\tilde {\bf C}) + \psi_3 (\tilde {\bf C}) 
$
where
$\psi_1  =  {I_1}^4\,I_2 - \frac{41\,{I_2}^3}{99} +
\frac{7\,I_1\,I_2\,I_3}{66} + \frac{{I_3}sof^2}{1056}$ and $                                           
\psi_2  =   \frac{4\,{I_2}^3}{11} + {I_1}^3\,I_3 - \frac{8\,I_1\,I_2\,I_3}{11} + \frac{17\,{I_3}^2}{528}.
$
The value of parameter $\beta_1=-0.25$ ($\beta_1= 4$) must be set    to ensure that the global minimum of the energy corresponds to square (triangular)  symmetry.
 The proposed energy $\tilde \phi_D( \tilde {\bf C}) $ concerns metrics located  on the surface $\det{\bf C}=1$. To account for configurations which also allows for a volume change,  we can add a  volumetric term to $\tilde \phi_D( \tilde {\bf C}) $.  For instance, to exclude   configurations with infinite compression one can use  an expression
$
  h(\det{\bf C}) =  -K (\ln\det {\bf C}-\det {\bf C}),
$
so that   $\phi_D( {\bf C})=  \tilde\phi_D(\tilde{\bf C}) + h(\det{\bf C}) $ where the coefficient K plays the role of a bulk modulus.   The energy density $\phi_D( {\bf C})$ is used in all numerical experiments reported in this paper. 

\paragraph{Internal length scale.} Since  the energy $\phi$ is  non convex, the corresponding continuum  elasticity problem,  which is by definition scale free,  is highly degenerate. 
 The minimization in this setting can produce  infinitely fine microstructures \cite{Ortiz1998-vi} reducing the stiffness in the relaxed problem  to zero \cite{fonseca1987variational}.  This lack of convexity is a property that the MTM of crystal plasticity shares with other similar Landau type theories.  However, in contrast to the conventional Ginzburg-Landau approaches,   relying for regularization on higher gradients of the order parameters,  in    MTM the  regularization  is achieved by spatial discretization, which  reduces the space of admissible deformations to a finite dimensional set of  compatible, piece-wise affine mappings. In other words, deformation is assumed to be piecewise linear and the    elastic  response is attributed   to   discrete material elements  whose scale $h$  defines the resolution of the model (meso-scale)  and is  viewed as a  physical parameter  \cite{Salman2011-jd,Baggio2019-rs}. 
 
 More specifically, the original lattice is coarse grained  with an introduction of a uniform meso-scale  grid reproducing the symmetry  of the crystal. The   scale  of the elements of the grid is selected  to make sure that  the Cauchy-Born type   energy  \cite{ericksen2005cauchy,Ericksen2008-kx}, computed by \textit{ab initio} methods for elements in the corresponding range of sizes,  is essentially   periodic in the interesting range of  strains. In many crystals  the  periodicity  at the level of the few first energy wells  can be  captured   already for $h \sim 10a$ where $a$ is the atomic scale.  In the resulting coarse grained  description, some microscopic features like, for instance, dislocation cores will emerge as    blurred  because the scales smaller than $h$ are effectively homogenized out. While  some aspects of a truly atomistic description will be then necessarily lost, for instance, the implied  cut-offs may compromise the short-range interaction of dislocation cores during dislocation reactions, the crucial meso-scopic  interactions at distances of the order and  larger than $h$  are expected to be  captured correctly.  If we normalize the linear size of the macroscopic sample  by setting  $L=1$,  we acquire  a small dimensionless  parameter $h/L=1/N$,  where $N^2$ is the number of the nodes in the mesoscopic finite-element  grid. For instance, if $h$ is  in $nm$ size range, the simulations with $N \sim 10^3$ would describe a micrometer size samples.  

\paragraph{Computational approach.} Solution of a continuum   elastic  problem implies  local  minimization of  the energy  $W=\int_{\Omega}\phi(\nabla\bold y ) d\bold x$ which is prescribed  on  a  reference domain  $\Omega$. We assume that the system is  loaded  by  an affine displacement field prescribed  on $\partial \Omega$ (hard device).   The conditions of mechanical equilibrium  read $ 
\nabla\cdot\mathbf{P}=0,   
$ 
where ${\bf P}= \partial\phi/{\partial\bf F} $ is the   Piola-Kirchhoff stress tensor. 
Using the Eulerian   $i,j=1,2$ and the Lagrangian  $K,L =1,2$ indexes and assuming summation on repeated indexes, we can rewrite  the  equations  in the form
$ 
A_{iKjL} y_{j,KL}=0,
$ 
where $A_{iKjL}$ is the tensor of the tangential elastic moduli:
$A_{iKjL}= \frac{\partial^2\phi^0{({\bf C}^0 )}}{\partial F_{iK}\partial F_{jL}}.
$ Here  $ {\bf C}^0=\bold m^T\bold C\bold m$, where the integer-valued matrix $\bold m$  can be computed for each value of $\bold C$ using the Lagrange reduction algorithm. 

\begin{figure}[h!]
     \centering
     \begin{subfigure}{0.43\textwidth}
         \centering
         \includegraphics[width=\textwidth]{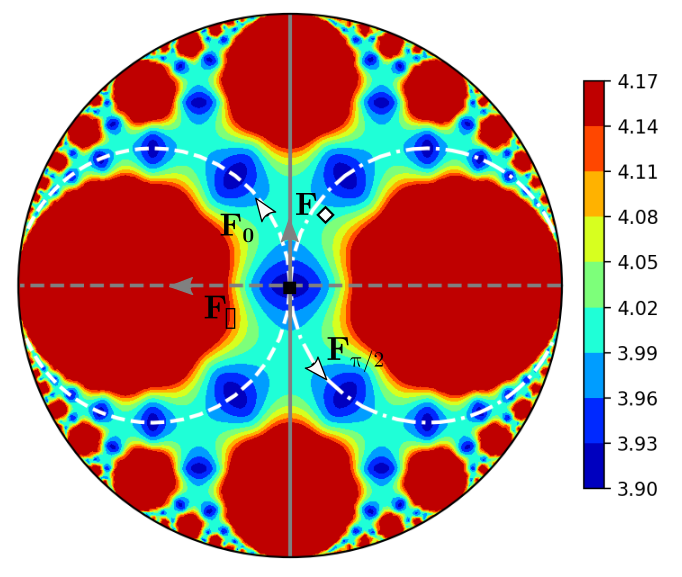}
         \caption{}
          \label{fig:en_sq_a}
     \end{subfigure}
     \begin{subfigure}{0.4\textwidth}
         \centering
         \includegraphics[width=\textwidth]{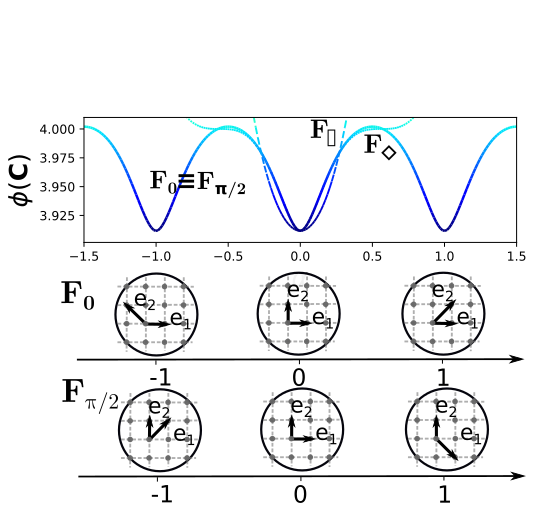}
         \caption{}
       \label{fig:en_sq_b}
     \end{subfigure}
\caption{\footnotesize{{\bf (a)} Poincaré disk  showing the energy landscape for the case  $\beta=-1/4$. Minima of the energy are located on square lattice configurations. Loading paths corresponding to simple shears are illustrated in white (dashed for ${\bf F}(\alpha, 0)$ and dash-dotted for ${\bf F}(\alpha, \pi/2)$. Grey lines are the rhombic pure shear ${\bf F}_{\diamond}$ (solid) and the rectangular ${\bf F}_\square$ (dashed). {\bf (b)} Energy landscape along the loading paths ${\bf F}(\alpha, 0)$ and ${\bf F}(\alpha, \pi/2)$, the equivalent  configurations  of the square lattice are   illustrated on the insets below the corresponding energy  wells. Energy landscape along the rectangular and the rhombic pure shear paths ${\bf F}_\square$ and ${\bf F}_{\diamond}$ are also shown for comparison. }} \label{fig:en_sq}
\end{figure}

The meso-scopic finite elelment grid is formed by a network of nodes, labelled by  integer valued coordinates $a =1,..., N^2$. We assume that each   element of the network is a deformable triangle  and write  the displacement field   in the form ${\bf u} ({ \bf x})= {\bf u}^a {\mathcal N}^a(\bf x )$, where ${\mathcal N}^a(\bf x)$ are  the  compactly supported  shape functions,  ${\bf u}^a$ are the amplitudes of nodal displacements and summation over repeated indexes effectively extends over elements containing or  bounding point $\bf x$. 
%
%
%
The  mesoscopic deformation gradient is   then    ${\bf F}(\bf x)  = \mathbb{I}  +\nabla {\bf u}(\bf x)$, 
and the 
 equilibrium equations can be written in the form
$
 \partial W/\partial {\bf u}^a  =\int _{\Omega} {\bf P}({\bf F}) \nabla {\mathcal N}^a ({\bf x}) d  {\bf x}  =0.
$
The hard device loading is set through the displacement ${\bf u}(\alpha)= (\bar{\bf F}(\alpha)-\mathbb{I}){\bf x}$ for all nodes $a$ on the boundary of the body $\partial \Omega$, where $\bar{\bf F}(\alpha)$ is the applied deformation gradient with amplitude $\alpha$. We also performed simulations with periodic boundary conditions ${\bf u}^B - {\bf u}^A= (\bar{\bf F}(\alpha)-\mathbb{I})({\bf x}^B-{\bf x}^A)$, where $A$ and $B$ are two points periodically located on the boundary of the body $\partial \Omega$. The equilibrium problem   can be solved by quasi-Newton  method followed by the  so called  NR `refinement'  when  the initial guess is too far from the solution for Newton–Raphson method  to converge initially~\cite{Salman2021-sn}. 

More specifically, to find  ${\bf u}^a$ we first   use  the L-BFGS algorithm~\cite{Bochkanov2013-lk} which      builds a positive definite linear approximation   allowing one  to make a quasi-Newton step lowering  $W$.  Such  iterations continue till  the increment  of  total energy $W$ becomes sufficiently small. The obtained approximate solution is then   used as an initial guess  ${\bf w}^a$ to solve, using LU factorization~\cite{Sanderson2016-ht},  the  equations for the correction  $\text{d}{\bf w}^a$ which read
$
 K^{ab}_{ij}dw_j^b+R_i^a =0,
$
where  
$
K^{ab}_{ij}= A_{iKjL}({\bf F}) \frac{\partial {\mathcal N}^a}{\partial x_K}\frac{\partial {\mathcal N}^b}{\partial x_L}$ and $
R^a_i=  P_{iK}( {\bf F}) \frac{\partial {\mathcal N}^a}{\partial x_K}.
$
 The displacement field can be updated in this way till   the value of  the forces acting on the nodal points are sufficiently small and  then  the loading parameter can be  advanced again, see \textbf{Appendix 1} for more details.

\section{Loading paths } 
\label{sec:staban}
\begin{figure}[h!]
     \centering
     \begin{subfigure}{0.43\textwidth}
         \centering
         \includegraphics[width=\textwidth]{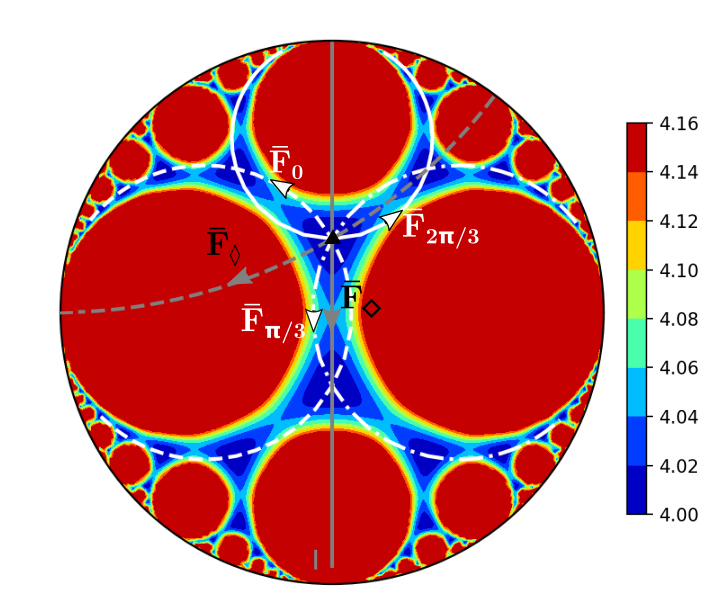}
         \caption{}
          \label{fig:en_hex_a}
     \end{subfigure}
     \begin{subfigure}{0.35\textwidth}
         \centering
         \includegraphics[width=\textwidth]{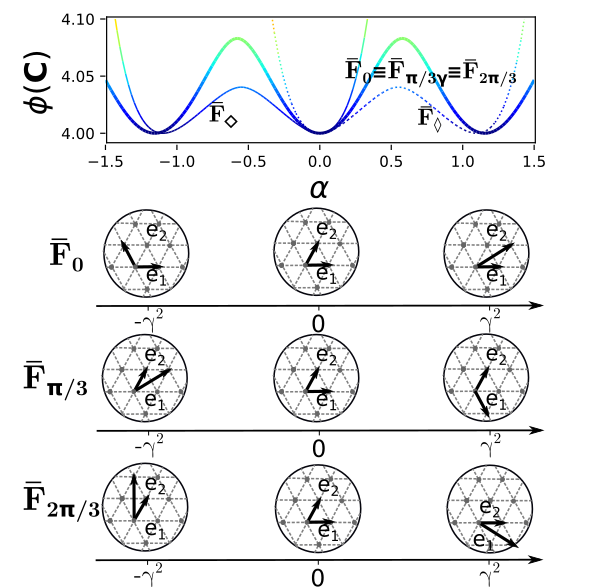}
         \caption{}
       \label{fig:en_hex_b}
     \end{subfigure}\caption{\footnotesize{{\bf (a)} Poincaré disk  showing the energy landscape for the case $\beta=4$. Minima of the energy are located on triangular lattice configurations and simple shears form circular trajectories (shown in white). The loading paths   ${\bf F}_\diamond$ and   $\bar{\bf F}_\lozenge$   are illustrated in grey (with a continuous and a dashed line respectively). {\bf (b)} Energy landscape along shearing deformation paths  $\bar{\bf F}(\alpha, 0)$, $\bar{\bf F}(\alpha, \pi/3)$ and  $\bar{\bf F}(\alpha, 2\pi/3)$, the shear-invariant triangular configurations are   illustrated below the corresponding energy  wells. The non-symmetric energy landscapes along the two pure shear paths are shown as well for comparison.  
}} \label{fig:en_hex}
\end{figure}

In Figs. \ref{fig:en_sq} and \ref{fig:en_hex}, we illustrate the energy landscapes in the cases when either   square and a triangular lattice is chosen as the ground state. While some details are specific of the polynomial form of the energy density chosen in this work (say,  the  size of  energy barriers) most of the observed features are generic and directly related to the symmetry requirements imposed on the energy. To illustrate   the periodic nature of such energy we show in the insets its evolution along selected shearing deformation paths. 

\paragraph{Square lattice.} Consider first the case of a lattice with square symmetry. Slip systems correspond in this case to the simple shear trajectories described by deformation gradients  of the type \begin{equation}\mathbf{F}(\alpha,\theta)=\mathbf{I}+\alpha\mathbf{R}(\theta)\mathbf{e}^0_1\otimes \mathbf{R}(\theta)\mathbf{e}^0_2,\label{eq:shear}\end{equation} where   $\mathbf{e}^0_i$ are the vectors of the reference orthonormal basis, $\mathbf{R}(\theta)$ is an orthogonal matrix representing a clockwise rotation at the angle  $\theta$ with respect to  $\mathbf{e}^0_1$ and $\alpha$ is the shear amplitude parameter.  The associated strain tensors ${\bf C}$ follow  circular trajectories on the Poincar\'e disk. In Fig. \ref{fig:en_sq} the white continuous and dotted circles correspond respectively to shears ${\bf F}(\alpha,\theta=0)$ and ${\bf F}(\alpha,\theta=\pi/2)$, which are oriented along close packed directions.
In Fig. \ref{fig:en_sq}(b), we illustrate the energy landscape along such simple shear trajectories with the corresponding deformed lattice configurations  shown  below. 
%

While both 'soft' and 'hard' simple  shear  loading paths were considered in detail in \cite{Salman2021-sn}, in this paper we focus  on the pure shear paths, that is, on volume preserving deformations that shrink the elementary  cell of the crystal along one  axis while elongating it along another one which is oriented in the perpendicular direction.   We consider two pure shear loading paths for which the  corresponding  metric tensors $\bf C$ are non-generic as they are located on the boundaries of the fundamental domain $\bf D$.
In the purely elastic regime such loading protocols transform   the original square configurations  into  either rectangular and rhombic loaded configurations without changing their specific  volumes; in what follows we use the notation $\mathbf{F}_\diamond$  for the rhombic pure shear  and $\mathbf{F}_\square$ for the  rectangular pure shear. 

Along the rhombic path the direction $ -(\sqrt{2}/2){\bf e}_1+(\sqrt{2}/2){\bf e}_2$ is shortened  while the direction  $ (\sqrt{2}/2){\bf e}_1+(\sqrt{2}/2){\bf e}_2$ is elongated with the volume of the element remaining constant. Then, 
${\bf C}_\diamond={\bf F}_\diamond^{T}{\bf F}_\diamond={\bf U}_\diamond^{T}{\bf R}^{T}{\bf R}{\bf U}_\diamond={\bf U}_\diamond^{2}
$, where 
\begin{align}
{\bf U}_\diamond&=\boldsymbol{\Psi}\boldsymbol{\Lambda}^{1/2}\boldsymbol{\Psi}^{T}=\left[\begin{array}{cc}
\frac{\sqrt{2}}{2} & \frac{\sqrt{2}}{2}\\
\frac{-\sqrt{2}}{2} & \frac{\sqrt{2}}{2}
\end{array}\right]\left[\begin{array}{cc}
\frac{1}{\lambda} & 0\\
0 & \lambda
\end{array}\right]\left[\begin{array}{cc}
\frac{\sqrt{2}}{2} & -\frac{\sqrt{2}}{2}\\
\frac{\sqrt{2}}{2} & \frac{\sqrt{2}}{2}
\end{array}\right]
\label{eq:pure_shear1}
\end{align}
is the  the stretch tensor,  $\boldsymbol{\Psi}$ is the orthogonal matrix whose columns are the principal directions and $\boldsymbol{\Lambda}$ is the diagonal matrix  with the squares principal stretches $\lambda_i$ as eigenvalues \cite{Thiel2019-wy}.  
The corresponding deformation gradient, chosen in such a way that  the lower side of the element is  aligned with the horizontal direction during the  deformation process, can be  written as $\mathbf{F}_\diamond=\mathbf{R}_\diamond\mathbf{U}_\diamond$, where 
\begin{equation}
\mathbf{F}_\diamond=\frac{1}{\sqrt{\cosh\alpha}}\left[\begin{array}{cc}
\cosh\alpha & \sinh\alpha\\
0 & 1
\end{array}\right]\, , \, \mathbf{R}_\diamond=\frac{1}{\sqrt{\cosh\alpha}}\left[\begin{array}{cc}
\cosh(\alpha/2) & \sinh(\alpha/2)\\
-\sinh(\alpha/2) & \cosh(\alpha/2)
\end{array}\right]
 \,,\label{eq:rhombic_path}
\end{equation}
and $\alpha=\ln{\lambda}$. We note that the rhombic   path is tangent to the simple shear path  $F(\alpha,0)={\bf I}+\alpha{\bf e}_1^0\otimes{\bf e}^0_2$,  these two deformation directions  are   interchangeable  in the classical linear elasticity (but not in MTM).


%
%
%

Along the rectangular path $\mathbf{F}_\square$ the principal directions are the reference vectors $\mathbf{e}^0_1$ and $\mathbf{e}^0_2$, therefore:
\begin{equation}
\mathbf{F}_\square={\bf U}_\square =\left[\begin{array}{cc}
\frac{1}{\lambda}\,\,   0\\
0 \,\,  \lambda
\end{array}\right]=
 \left[\begin{array}{cc}
\cosh(\frac{\alpha}{2})-\sinh(\frac{\alpha}{2}) \,\,\,\,\,\,\,\,\,\,\,\,\,\,\,\,  0\\
0 \,\,\,\,\,\,\,\,\,\, \,\,\,\,\,\,  \cosh(\frac{\alpha}{2})+\sinh(\frac{\alpha}{2})
\end{array}\right]\,.
\label{eq:rectang_path}
\end{equation} 
The individual  elements are then elongated along the horizontal direction $\mathbf{e}^0_2$ and shortened along the vertical direction  $\mathbf{e}^0_1$. 
\begin{figure}[h!]
     \centering
     \begin{subfigure}{0.35\textwidth}
         \centering
         \includegraphics[width=\textwidth]{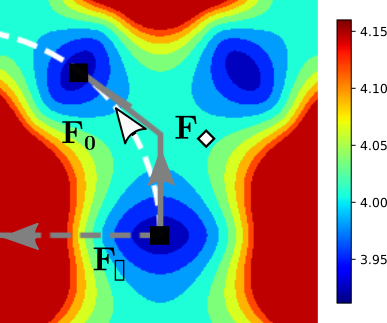}
         \caption{}
          \label{fig:en_sq_saddle_a}
     \end{subfigure}
\hspace{0.05\textwidth}
     \begin{subfigure}{0.45\textwidth}
         \centering
         \includegraphics[width=\textwidth]{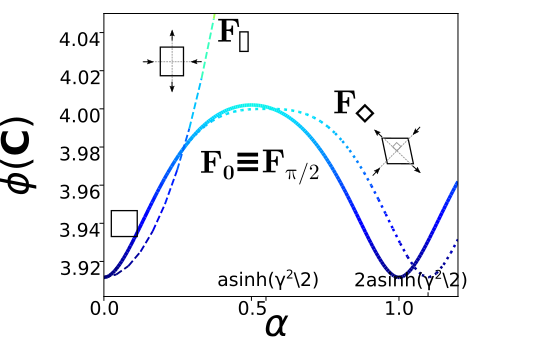}
         \caption{}
       \label{fig:en_sq_saddle_b}
     \end{subfigure}
\caption{\footnotesize{{\bf (a)} Loading paths $\mathbf{F}_{0}$ (dashed white) ,  ${\bf F}_\square$ (dashed grey) and  ${\bf F}_\diamond$ (grey). The latter two correspond to the boundary of the periodicity domain $D$ and describe the deformations of the square lattice  towards rectangular and rhombic configurations, respectively.  {\bf (b)} Energy landscape along the illustrated paths. The low energy path ${\bf F}_\diamond$  spans the bottom of the energy barrier and crosses the high symmetry point $\bf T$. }} \label{fig:en_sq_saddle}
\end{figure}

In  Fig. \ref{fig:en_sq_saddle}(a), we  the rhombic and the rectangular pure shear  loading paths superimposed on the energy surface  of a square crystal. One can see that the   rhombic path is located inside the energy valley and can be then considered as   'soft'. Instead, the  rectangular path goes against a steep energy hill and is therefore 'hard'. The corresponding one-dimensional energy landscapes are illustrated in Fig. \ref{fig:en_sq}(b).

\paragraph{Triangular lattice.} We now consider as the reference state,  where the loading path begins, the triangular lattice  $\bf T$.   Its generating  basis  is given by the two vectors ${\bf e}^{\triangle}_1=\gamma\left\lbrace 1,0\right\rbrace^T$ and ${\bf e}^{\triangle}_2=\gamma\left\lbrace 1/2 , \sqrt{3}/2 \right\rbrace^T$, with $\gamma=\sqrt[4]{4/3}$. The  shear paths are now characterized by the families of deformation gradients  
 \begin{equation}\bar{\bf F}(\alpha,\theta)=\mathbf{F}(\alpha,\theta){\bf H},\label{path:tri}\end{equation} where ${\bf H}$ is the matrix whose columns are the   basis vectors ${\bf e}^\triangle_i$, ${\bf F}(\alpha,\theta)$ is   shear deformation defined in Eq. \ref{eq:shear} for simple shears (we recover closed-pack directions for  $\theta=0,\pi/3,2\pi/3$). Note that with this parametrization,  the value of the parameter $\alpha$ for which the  lattice invariant shears for triangular symmetry are recovered is not an integer, but  instead $\alpha=n\gamma^2$ where $n$ is integer.  The energy profile along  these   paths    $ \bar{\bf F}(\alpha,\theta)$ is   shown  in Fig.~\ref{fig:en_hex}(b), see  \cite{Salman2021-sn} for more details.
 
\begin{figure*}[h!]
\centering
\includegraphics[width=0.8\textwidth]{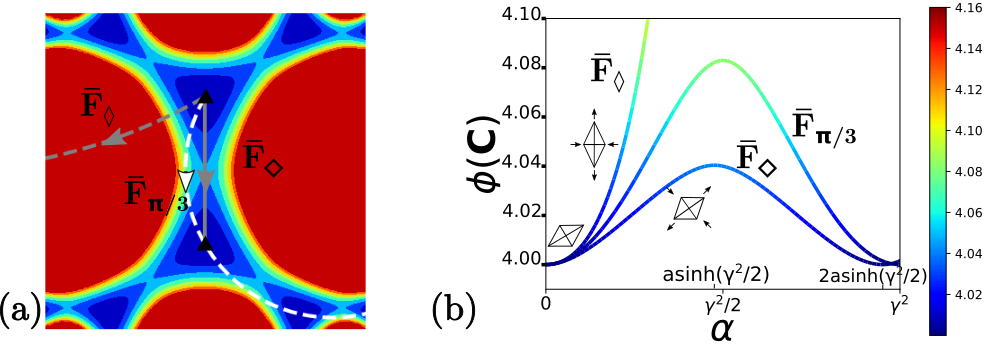}
\caption{\footnotesize{{\bf (a)} Loading path $\bar{\bf F}_{\pi/3}$ (dashed white), $\bar{\bf F}_\lozenge$ (dashed grey) and  $\bar{\bf F}_\diamond$ (grey). The latter two correspond  to the boundary of $D$ and describe the deformation of the triangular lattice towards different  rhombic configurations.   {\bf (b)} Energy landscape along the illustrated paths. The low energy path $\bar{\bf F}_\diamond$  spans the bottom of the energy barrier and crosses the high symmetry point $\bf S$. }} \label{fig:en_hex_saddle}
\end{figure*}

Here we focus instead on pure shear loading paths originating in triangular reference state $\bf T$ and  corresponding to the boundaries of the minimal periodicity domain $\bf D$. Along one of these paths,  $\bar{\bf F}_\diamond$, we obtain lattices with rhombic symmetry  where both diagonals of the rhombus are longer than the side; the other path, $\bar{\bf F}_\lozenge$, corresponds to the case of rhombi with  one of the diagonals   smaller than the side \cite{Conti2004-sv}.  We remark that the path $\bar{\bf F}_\diamond$  originating in $\bf T$ describes  the same deformation as the path ${\bf F}_\diamond$  originating  in  $\bf S$. In the case of triangular lattice, the principal directions  are rotated by  $\pi/6$ with respect to the reference axes of the square lattice, therefore, in analogy with (\ref{eq:pure_shear1}) one can write
  \begin{equation}
\bar{\bf U}_\diamond =\bar{\boldsymbol{\Psi}}\boldsymbol{\Lambda}^{1/2}\bar{\boldsymbol{\Psi}}^{T}\nonumber  
 =\left[\begin{array}{cc}
\cosh(\frac{\alpha}{2})-\frac{1}{2}\sinh(\frac{\alpha}{2}) & -\frac{\sqrt{3}}{2}\sinh(\frac{\alpha}{2})\\
-\frac{\sqrt{3}}{2}\sinh(\frac{\alpha}{2}) &  \cosh(\frac{\alpha}{2})+\frac{1}{2}\sinh(\frac{\alpha}{2})\\
\end{array}\right].
\label{eq:pure_shear_fat}
\end{equation} 
Among all such deformations   the one which preserves the angle between ${\bf e}^{\triangle}_1$ and the horizontal direction is $\bar{\mathbf{F}}_\diamond={\bf R}(\chi)\bar{\bf U}_\diamond$, with:
$
\chi=\arctan \left( \frac{\sqrt{3}}{2} \frac{\tanh{\alpha/2}+2}{2\tanh{\alpha/2}} \right)^{-1}.
$
 This deformation is then applied to the triangular basis ${\bf e}^\triangle_i$. Note that along the loading path ${\bf F}_\diamond$, the crystal   is driven  trough a very shallow energy valley extending from the (triangular) energy minimum $\bf T$ towards the mountain pass represented by the (square) saddle $\bf S$ and then further to another energy (square) minimum at $\alpha=2\arcsinh(\gamma^2/2)$ (see Fig. \ref{fig:en_hex_saddle}).  We remark that, along the 'soft' pure shear path $\bar{\bf F}_\diamond$, the energy barrier, which has its maximum at $\mathbf{S}$ (with $\alpha=\text{arccosh}(\gamma^2)$), is lower than the one along the simple shear path ${\bf F}_{\pi /3}$, the one which is  habitually selected as the natural 'plastic mechanism'.

 The second  rhombic loading  path $\bar{\bf F}_\lozenge$ is obtained   by applying the pure shear deformation ${\bf F}_\square$ to the lattice defined by the basis vectors ${\bf e}^\triangle_i$. Along  the path $\bar{\bf F}_\lozenge$  which is much 'harder' than the path ${\bf F}_\diamond$, the energy grows very rapidly   without ever passing through  any other minimum, see Figure \ref{fig:en_hex_saddle} .

 %

%

%
%

\paragraph{Stability limits.}
With each loading path we can associate an effective stability (yield) limit obtained under the assumption that the state is homogeneous and the discretization length scale is vanishingly small. In other words, we imply here an instability of a perfect crystal deformed  in a hard device with the affine deformation  $\bar{\bf F}(\alpha)$ applied on the boundary and search for the critical value of the loading parameter $\alpha_c$ at which the homogeneous state ceases to be stable. To identify the bifurcation point we need to solve an   incremental problem defined by the tangential  elastic moduli 
$
\mathcal{A}_{iKjL}=\frac{\partial^2\phi}{\partial F_{iK}\partial F_{jL}}
$
It is known that 
%
the homogeneous configuration   remains  incrementally stable in the above sense as long as the Legandre-Hadamard (strong ellipticity condition) 
$
Q_{ij}(\mathbf{N})l_i l_j>0
$
holds \cite{ogden1997non}, where  we introduced the  acoustic tensor $Q_{ij}(\mathbf{N})= \mathcal{A}_{iKjL} N_K N_{L}$ while  $\mathbf{N}$ and $\mathbf{l}$ are arbitrary  vectors, in the reference and deformed configurations, respectively. The corresponding critical value of the loading parameter   is  usually found from the condition  $\det Q(\mathbf{N})=0$, e.g.  \cite{Borja2001-hr}. In what  follows  we  use an Eulerian version of this bifurcation condition $\det \mathbf{q}(\mathbf{n})=0$, where $q_{ik}=\mathbf{a}_{ijkl}n_jn_l$, $\mathbf{a}_{ijkl}=\mathcal{A}_{iKjL}F_{kK}F_{lL}$ and  $\mathbf{n}=\mathbf{F}^{-T}\mathbf{N}$. The Eulerian vectors $\bf n$ and $\bf l$  characterize the incipient unstability mode \cite{rice1976localization}. For instance,  if $\bf n$ is approximately perpendicular to $\bf l$. In the post-bifurcational regime  one can  expect in this case the formation of  (lattice size)  shear bands along the  plane with normal  $\bf n$ and with slip direction $\bf l$ \cite{Van_Vliet2003-yg}.  Further development may lead to the  nucleation inside the  individual  bands  of incipient dislocation pairs (slip embryos in 2D or dislocation loops in 3D)  whose Burgers vector is  aligned  with $\bf l$ or to the collective process resulting in activation of a micro-twin laminate  with the twinning plane oriented along $\bf n$.
\begin{figure}[t!]
     \centering
     \begin{subfigure}{0.25\textwidth}
         \centering
         \includegraphics[width=\textwidth]{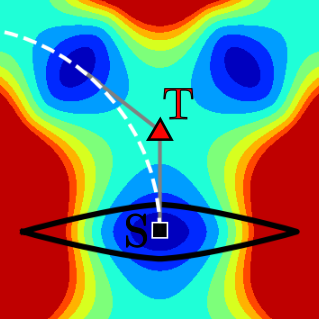}
         \caption{}
         \label{fig:yield_surf_sq}
     \end{subfigure}
\hspace{0.1\textwidth}
     \begin{subfigure}{0.25\textwidth}
         \centering
         \includegraphics[width=\textwidth]{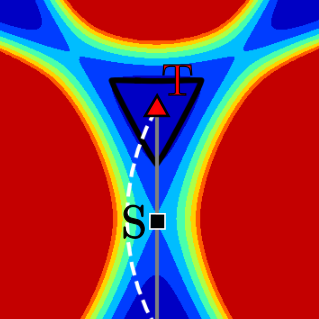}
         \caption{}
        \label{fig:yield_surf_hex}
     \end{subfigure}
     \caption{\footnotesize{{\bf (a)} Stability (yielding) limits  for the square crystal (in black). The two loading paths corresponding to simple shear ${\bf F}_0$ (dashed white) and rhombic pure shear ${\bf F}_\diamond$ (grey) cross the stability region in similar  configurations. {\bf (b)} Stability region for the triangular symmetry crystal (black). Here the difference of  strain configurations at the limit of  stability is larger if we  compare    the   loading paths  corresponding to  simple shear (dashed white) and  pure shear (grey).}}
\end{figure}

Using the proposed approximate  stability condition we can delineate  in the configurational space of metric tensors $\bf C$  a region around the reference state where  the continuous homogeneous system can be expected to be  stable and interpret it as an effective 'yield surface'. To this end we need to  consider   a sufficiently broad family of loading paths, for instance, the family of simple shear trajectories with the full range of  values  of the shearing angle $\theta$ plus the  two limiting loading paths along the boundary of the periodicity domain $\bf D$ and representing pure shears  (the paths ${\bf F}_\square$ and ${\bf F}_\diamond$ for the square lattice, and $\bar{\bf F}_\diamond$ and $\bar{\bf F}_\lozenge$ for the triangular lattice). Along each of these paths we   computed  the first value of the loading parameter $\alpha$ where the Legandre-Hadamard condition is violated for some   non-trivial $\bf n$ and $\bf l$. This   produced an effective 'yield surface'  which we  illustrated by black lines  in our Figures \ref{fig:yield_surf_sq} and \ref{fig:yield_surf_hex} for square and triangular lattices, respectively.

\begin{figure}[hbt!]
     \centering
     \begin{subfigure}{0.4\textwidth}
         \centering
         \includegraphics[width=\textwidth]{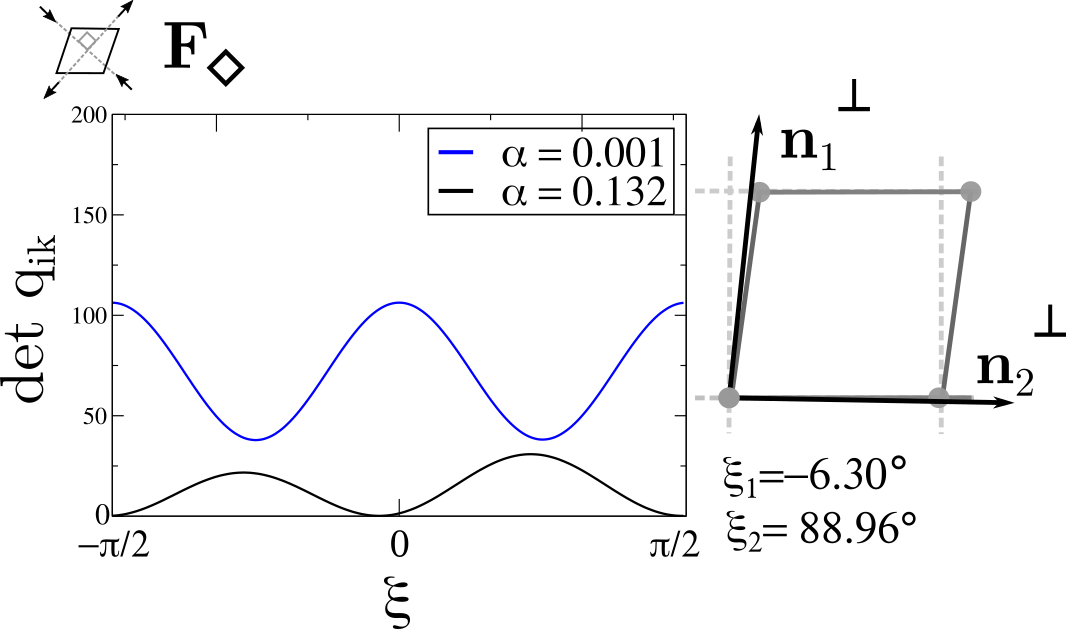}
         \caption{}
     \end{subfigure}
\hspace{0.01\textwidth}
     \begin{subfigure}{0.4\textwidth}
         \centering
         \includegraphics[width=\textwidth]{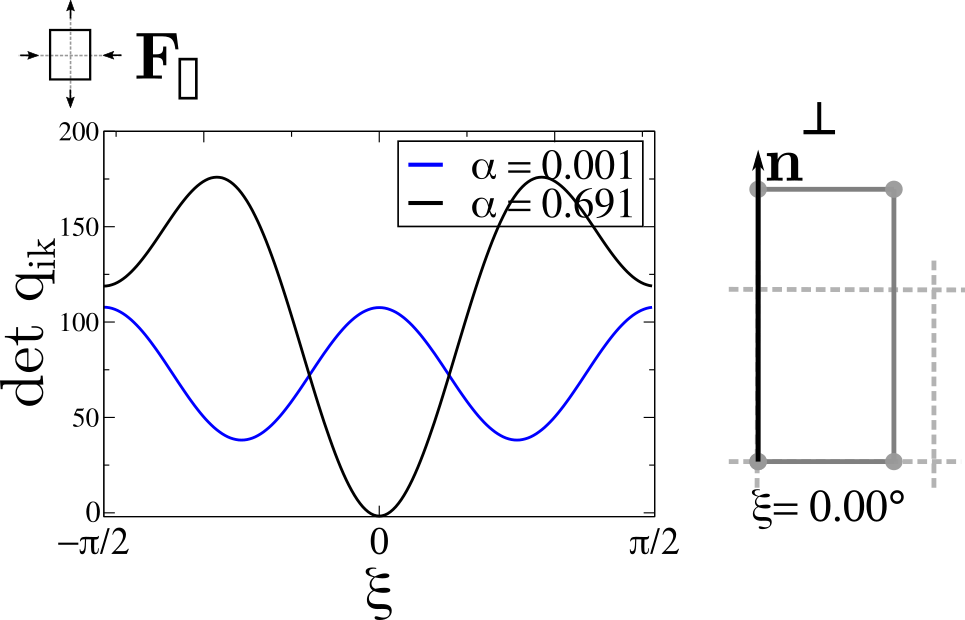}
         \caption{}
     \end{subfigure}
     \caption{\footnotesize{The function  $\det\left({\bf F}(\alpha),{\bf n}(\xi)\right)$ is illustrated for $\alpha=0.001$ and $\alpha=\alpha_c$ for the two deformation paths considered in the case of the square   crystal, that is ${\bf F}_\diamond$ {\bf (a)}  and ${ {\bf F}}_\square$ {\bf (b)}.}\label{fig:yield_surf_sq1d}
}
\end{figure}
\begin{figure}[hbt!]
     \centering
     \begin{subfigure}{0.4\textwidth}
         \centering
         \includegraphics[width=\textwidth]{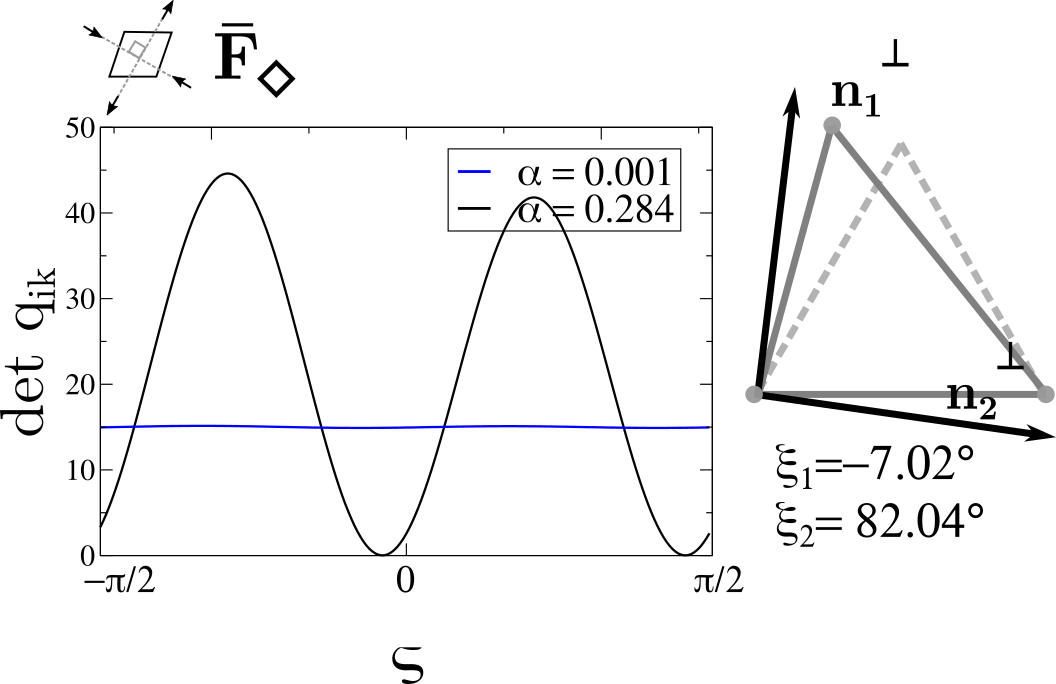}
         \caption{}
     \end{subfigure}
\hspace{0.01\textwidth}
     \begin{subfigure}{0.4\textwidth}
         \centering
         \includegraphics[width=\textwidth]{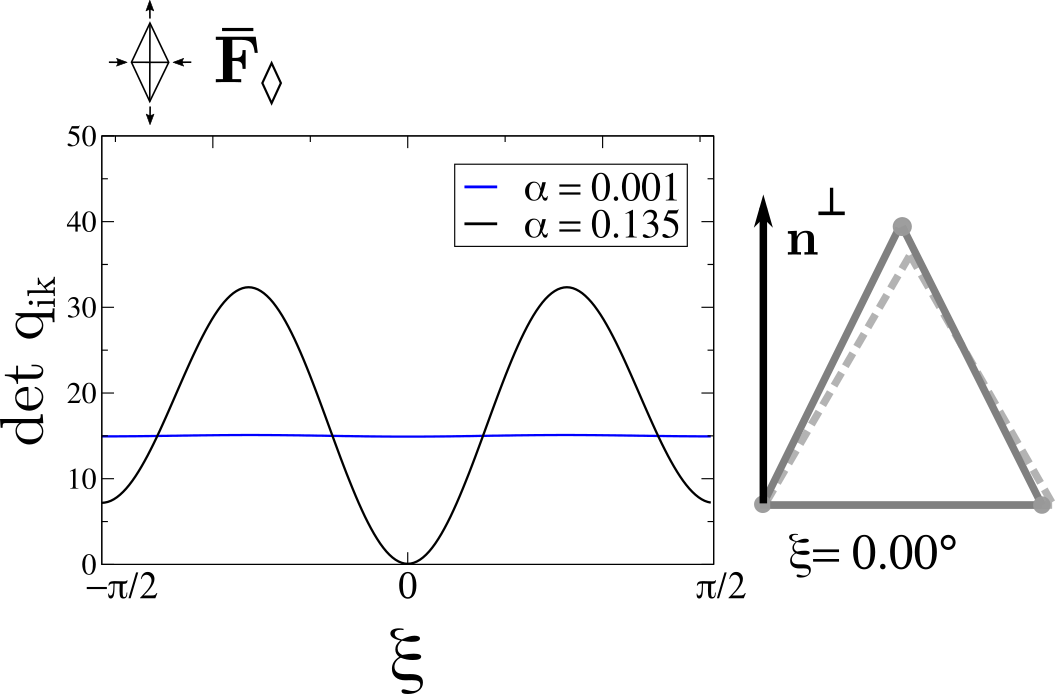}
         \caption{}
     \end{subfigure}
     \caption{\footnotesize{The function $\det\left({\bf F}(\alpha),{\bf n}(\xi)\right)$ is illustrated for $\alpha=0.001$ and $\alpha=\alpha_c$ for    two deformation paths  considered in the case of the triangular   crystal, that is $\bar{\bf F}_\diamond$ {\bf (a)} and ${\bar {\bf F}}_\lozenge$ {\bf (b)}.} \label{fig:yield_surf_hex2}}
\end{figure}

We illustrate  the nature  of the instability modes  for square and triangular lattices  loaded along the special pure shear  paths. If the potentially unstable orientation $\bf n$ is  parametrized by the  angle $\xi$   as $\bf n=\left\lbrace \cos\xi,\sin\xi \right\rbrace$ it is of interest to  study the $\xi$ dependence of the  parameter $\det \mathbf{q}(\mathbf{n}) $ at different values of  $\alpha$ and in our Fig. \ref{fig:yield_surf_sq1d} and Fig. \ref{fig:yield_surf_hex2}, we show such graphs for $\alpha=0$ and $\alpha_c$ for all four pure shear loading   paths discussed above. In the inset located to  the right of each of these plots we represented the directions ${\bf n}^\perp$ indicating the orientation of  the unstable (slip) plane vis a vis  the basis vectors of the deformed crystal   at the onset of instability ( along with the values of $\xi$). We note that for  the   rectangular path  ${\bf F}_{\square}$ for the square and $\bar{\bf F}_\lozenge$ for the triangular lattices the unstable mode is perfectly aligned with the horizontal plane (${\bf n}^\perp$ is aligned with the vertical directions).  The  polarization vectors $\bf l$ were found to be   approximately perpendicular to $\bf n$ for all of the investigated loading directions.


We remark that the simple shear type loading  paths were  discussed in detail in   \cite{Salman2021-sn} where we showed that for  square lattices  the  instability along the ('soft') simple shear direction ${F}_{\theta=0, \alpha}$ produces  two almost simultaneous instability modes  with the resulting  activation of two crystallographic slip systems. We have seen that such modes are also  almost  simultaneous in the case of square lattices subjected to the ('soft') pure shear loading ${\bf F}_\diamond$. Moreover, the analysis of the ('soft') path $\bar{\bf F}_\diamond$  for  triangular lattices shows the  analogous effect  ( which is  not apparent along crystallographic-oriented simple shears). 
Along the  generic ('hard') shearing directions (implying both pure and simple shears),  there is only one unstable mode $\bf n$  which reflects the activation of a single slip system.


 %
%

%

\section{Numerical experiments}\label{sec:simres}

In this section we present the results of our numerical simulations. Their main goal is to provide first  evidence of the efficiency of   MTM in addressing various  sub-continuum problems  in  crystal plasticity. Throughout this section we use the version of the model,   the numerical algorithm and the loading protocols  described in the previous sections.

\paragraph{Dislocation cores.} To interpret the obtained data in  experiments involving    large number of dislocations,  it is important to be able to identify and resolve the structure of individual dislocation cores. That is why we begin with consideration of an isolated dislocation   trapped by the discreteness of the meso-scopic lattice in the center of a sufficiently large unloaded crystal.

%
\begin{figure}[h!]
     \centering
     \begin{subfigure}{0.47\textwidth}
         \centering
         \includegraphics[width=\textwidth]{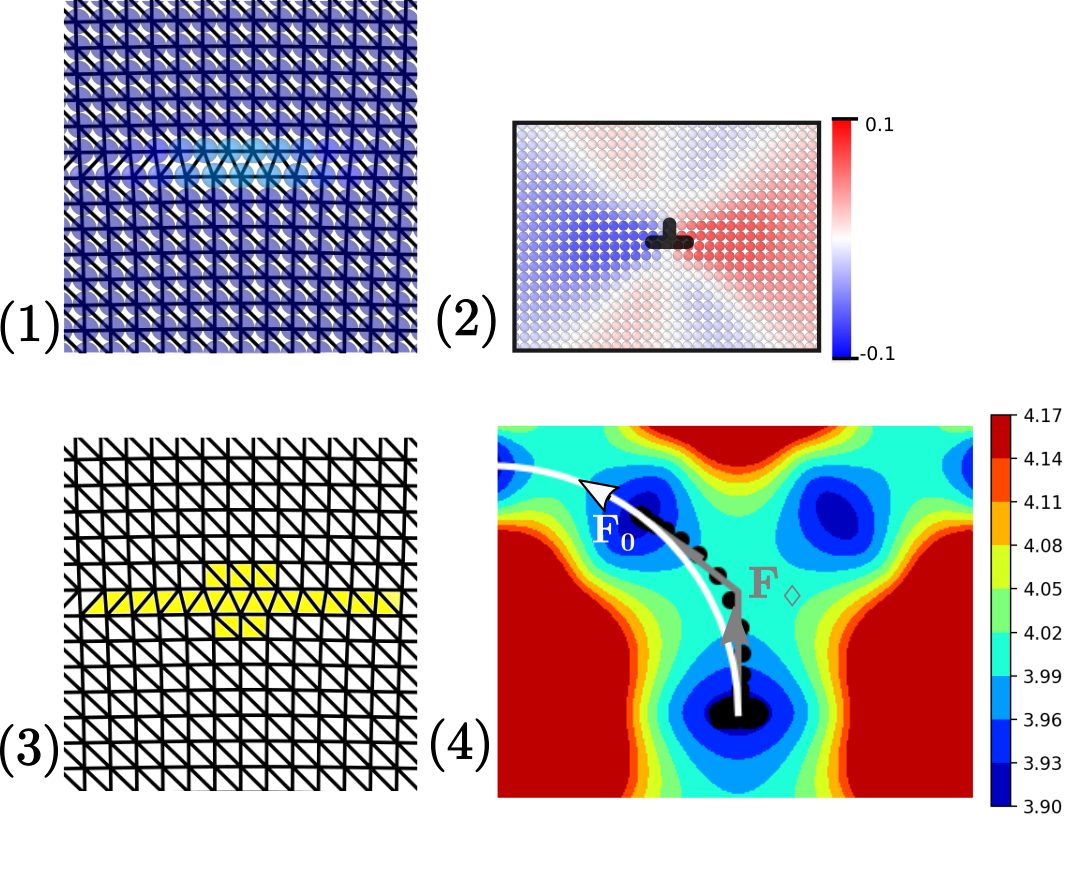}
         \caption{}
         \label{fig:dislo_sq}
     \end{subfigure}
\hspace{0.01\textwidth}
     \begin{subfigure}{0.47\textwidth}
         \centering
         \includegraphics[width=\textwidth]{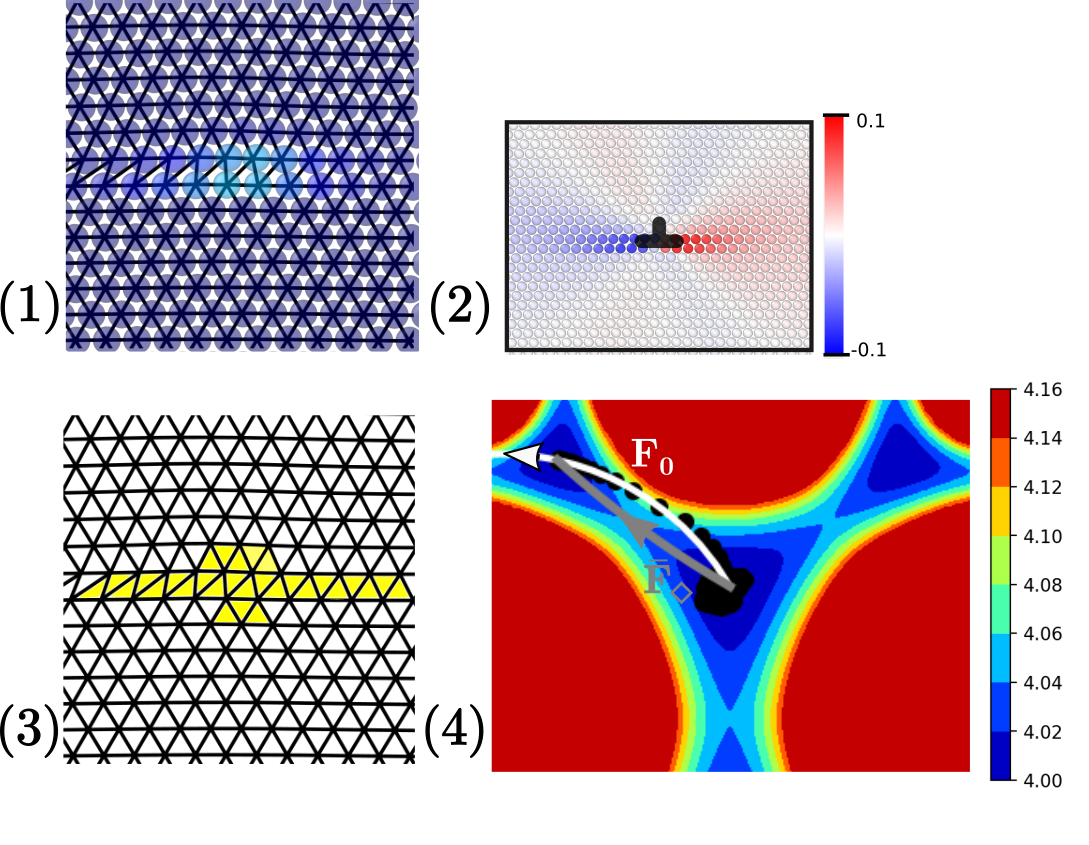}
         \caption{}
        \label{fig:dislo_hex}
     \end{subfigure}
\caption{\footnotesize{{\bf (a)} Dislocation structure in the case of square lattice: (1) energy near the core, (2) Cauchy stress $\sigma_{xy}$, (3) a detail of the elements triangulation, (4) elements strain projection on $\bf C$ space,  color bar shows the energy level  both in  {\bf (a)} and {\bf (b)}. {\bf (b)}  Dislocation structure for the triangular symmetry crystal. Pictures are analogous to {\bf (a)}. }} \label{fig:dislo}
\end{figure}
As we have already  mentioned,  dislocations can appear in MTM when different variants of the same lattice (different phases)  are present simultaneously. Consider,  for instance, the coexistence in the square lattice  of the   reference phase  ${\bf S}={\bf F}(\theta=0,\alpha=0)$ and the   phase ${\bf S}_{1}^{0}={\bf F}(\theta=0,\alpha=1)$ which is different from the reference phase by an elementary  lattice invariant shear.  A single  dislocation  is  obtained in the configuration where a semi-infinite  single  layer of elements in phase ${\bf S}_{1}^{0}$ is embedded in  an infinite lattice of elements in phase ${\bf S}$, see Fig. \ref{fig:dislo}(a). Far away from the area around the terminal point  of the sheared (slipped) layer of elements, which represents the dislocation core, the lattices are  perfectly compatible because all such elements lie in the bottoms of the corresponding energy wells. Elements in the core region lie outside the energy wells and have  therefore nonzero elastic energy.

 To obtain in a numerical experiment an isolated  dislocation  we used  a  square  domain (with  $200\times200$ finite element nodes)  and applied on its  boundary  the displacement field reproducing  anticipated far field continuum asymptotics (Volterra dislocation, \cite{Jaswon1991-js}), 
$
u_{x} =\frac{b}{2\pi}\left[\arctan\frac{y}{x}+\frac{xy}{2(1-\nu)(x^2+y^2)}\right]$ and 
$u_{y} =\frac{b}{2\pi}\left[\frac{1-2\nu}{4(1-\nu)}\ln(x^2+y^2)+\frac{x^2-y^2}{4(1-\nu)(x^2+y^2)}\right]$.
The  configuration of the nodes  was then allowed to  relax elastically till the local minimum of the energy was reached. As a result of such relaxation an isolated  dislocation core was formed in the middle of the  domain whose different representations (energy, stress, deformation) are shown in    Fig. \ref{fig:dislo}(a) for  the case of square lattice and   in Fig. \ref{fig:dislo}(b)  for the case of triangular lattice. In   Figs. \ref{fig:dislo} (a-4,b-4) we show the corresponding core structures in the configurational space of metric tensor.


From  the deformed configuration of the elastic elements shown  in Fig. \ref{fig:dislo}(a-3), one can see the  sheared layer to   the left  of the (square) dislocation core representing the (square) energy well ${\bf S}_{1}^{0}$ while the elements in the same layer but located on the right side   of the dislocation core are in reference (square) well ${\bf S}$.  Similarly, we see in Fig. \ref{fig:dislo}(b-3)  that the (triangular) dislocation core can be viewed as a domain boundary separating  the  coexisting    elements  of the two neighboring (triangular) energy wells ${\bf T}^{1}_{0}$ and $\bf T$.  The presence of all these energy wells    becomes  even more clear  as one  looks at the values of the components of the metric tensors $C_{11}, C_{12}, C_{22}$ at  the  elastic elements which allows one to represent the structure  of a core   as a (in reality, somewhat blurred) trajectory
in the configuration space, see our Fig.  \ref{fig:dislo}(a-4) and Fig. \ref{fig:dislo}(b-4). While the initial and the final points in such trajectories are located at the bottoms of the corresponding energy wells, the trajectories themselves represent a   mountain pass type connections between the wells. In the case of square crystals such trajectory  ensures that the maximal elevation is minimal but apparently, this is not the case for triangular crystals. This confirms that  while for both square and triangular lattices most of the transitions takes place  close to the bottoms of the energy valleys,  the   fine structure of the barriers  is  manifestly symmetry dependent.


Thus, in the case of the square lattice,  the trajectory describing   a dislocation core appear to consist of two separate segments (shown in grey in Fig.  \ref{fig:dislo}(a-4) representing    pure shears  of the type  ${\bf F}_\diamond$ studied in the previous section. Each of them connects the corresponding square wells (the reference well $\mathbf{S}$ and the equivalent well   $\mathbf{S}^{-1}_{\pi/2}=\mathbf{F}_{\pi/2}(\alpha=-1)$ reachable by an elementary lattice invariant shear) with the shallow local minimum (almost a monkey saddle for our choice of the potential, see \cite{Baggio2019-rs}) describing the triangular (hexagonal) lattice  $\bf T$. Here the configuration $\bf T$, whose  presence   in the core structure is also  suggested also by the configuration of the elements shown in  Figure \ref{fig:dislo} (1c),  plays here the role of a stacking fault   while the pure shears can be interpreted as the analogs of Shockley partials, see for instance  \cite{Kamimura2018-kz,Bulatov1997-vc,Mohammed2022-ed}. Note that the naively favored simple shear trajectory ${\bf F}_0$ (shown in white  in Figure  \ref{fig:dislo} (1-d))  delivers, as we have seen before,   a slightly  higher barrier   and is therefore avoided by the solution of the energy minimization problem.



 The structure of the dislocation core in triangular lattices  is different.  Thus, the corresponding mountain pass type trajectory in the configurational space (shown in white in Fig.  \ref{fig:dislo}(b-4))  follows the simple shear path ${\bf \bar{F}}_{0}$. An alternative trajectory consisting of two pure shear segments  and passing through the square energy configuration ${\bf S}$  (shown in gray  in Fig.  \ref{fig:dislo}(b-4)) is not taken by the system despite being characterized by a lower energy barrier (see Fig. \ref{fig:en_hex_saddle}).



%

\paragraph{ Collective nucleation of dislocations.} Now, instead of the specially designed non-affine  boundary conditions  ensuring  the emergence of a single dislocation, we consider   generic affine loading paths  and study the symmetry breaking decomposition of the  homogeneous state.  More specifically, we assume that the system is driven  quasi-statically and therefore evolves through a sequence of equilibrium configurations.  In the absence of pre-existing defects (pristine crystal), the initial  evolution of the system  from the unloaded reference state is elastic till the corresponding elastic branch of equilibria ceases to exist. At the point of instability the  dissipative   branch-switching event, accompanied by a macroscopic  stress, drop takes place.   It takes the form of  a system size  avalanche leading to collective  nucleation of a large number of dislocation and a global slip-induced reorganization of the crystal lattice.  

%

\begin{figure}[t!]
     \centering
     \begin{subfigure}{0.4\textwidth}
         \centering
         \includegraphics[width=\textwidth]{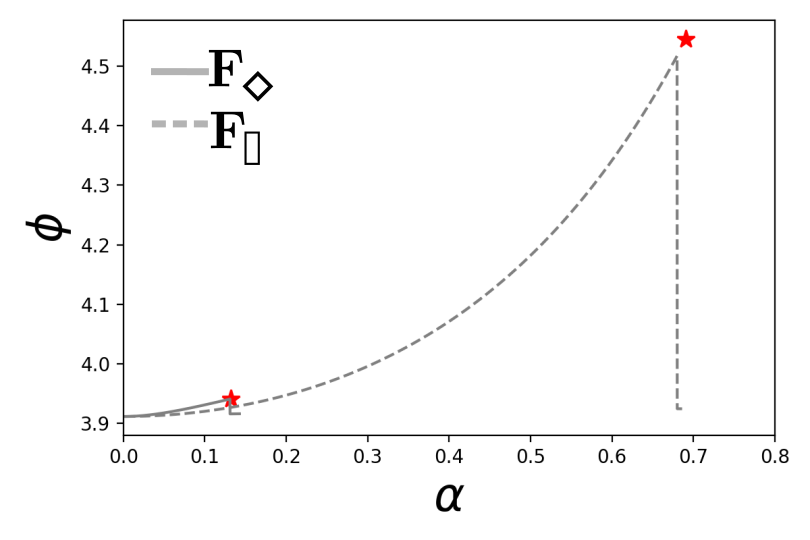}
         \caption{}
         \label{fig:inst_sq_fix}
     \end{subfigure}
\hspace{0.01\textwidth}
     \begin{subfigure}{0.4\textwidth}
         \centering
         \includegraphics[width=\textwidth]{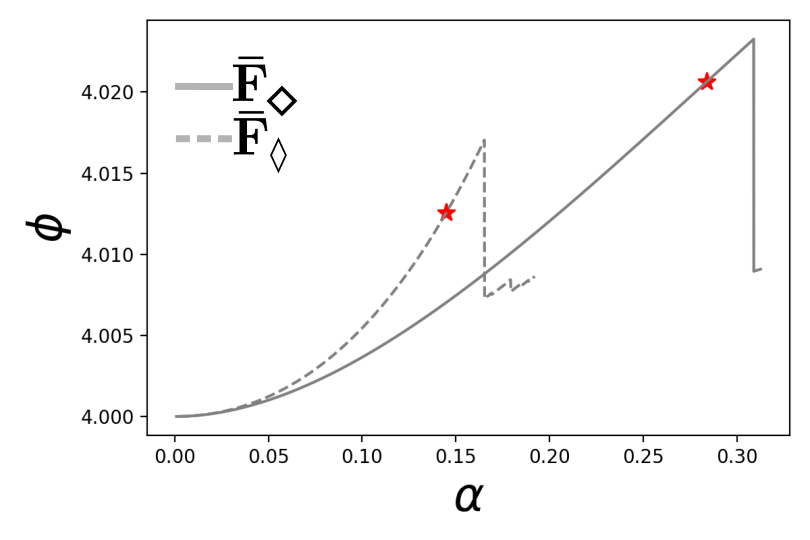}
         \caption{}
        \label{fig:inst_hex_fix}
     \end{subfigure}     
     \caption{\footnotesize{Energy-strain  relations obtained  in numerical experiments for a crystal with square symmetry: {\bf (a)}  and for a crystal of triangular symmetry {\bf (b)}. Simulated domains are formed by $N\times N$ nodes with $N=100$, loaded with affine  boundary conditions. Red stars display the critical parameters calculated analytically.}}
\end{figure}

Consider, for instance, the case  of a  square domain $ \Omega$ with  $N=100\times100$ nodes and assume that   the applied affine deformation is   a homogeneous simple shear $\bar{\bf F}(\alpha,\phi)$  with  fixed orientation $\phi$, and the shear amplitude   $\alpha$ playing the role of the loading parameter.  By changing this parameter in increments of $10^{-4}$,  we can advance   the  displacement field   ${\bf u}(\alpha,\phi)= (\bar{\bf F}(\alpha,\phi)-\mathbb{I}){\bf x}$ for all nodes  on the boundary of the body $\partial \Omega$ till the first instability  occurs signaling  the homogeneous dislocation nucleation.  The incremental solution  algorithm allowing one to see the unfolding of the avalanche in the fast computational time is   detailed in the  flowchart shown in Appendix \ref{alg:hudson}.  In \cite{Salman2021-sn} we  showed that the resulting (post-avalanche) dislocation pattern depends  on the orientation of the applied simple shear with a strong difference between the dislocational configurations  obtained in the cases of  soft and   hard loading directions. In the present  paper, we illustrate results obtained along the pure shear loading protocols using periodic boundary conditions  where the system   was loaded starting from a stable reference configuration till the point of instability close to the theoretically predicted elastic instability, see Fig. \ref{fig:inst_sq_fix} and Fig. \ref{fig:inst_hex_fix}. Results obtained along the same loading paths  but the  fixed boundary conditions are comparable in terms of observed collective dislocation mechanisms, but since they tend to display a stronger influence of the boundaries we are not  discussing  them here in detail. 

%
%
%


Here we report the results of the  numerical experiments obtained for the  pure shear loading protocols discussed in the previous section. These loading paths are of particular interest since they include  the 'softest' and the 'hardest' loading directions which correspond to  the 'shortest' and the  'longest' distance to instability', respectively. These loading paths are also highly symmetric which suggests that the post avalanche dislocation patterns  may have some particular features. Thus, as we have already seen, among the two pure shear loading directions, one is always directed towards the energy maximum and can be expected to produce regular micro-twin microstructures. Another one is aiming directly at the mountain pass where   the corresponding  saddle point may foment  the generation of disorder.

\begin{figure*}[h!]
\centering
\includegraphics[width=.7\textwidth]{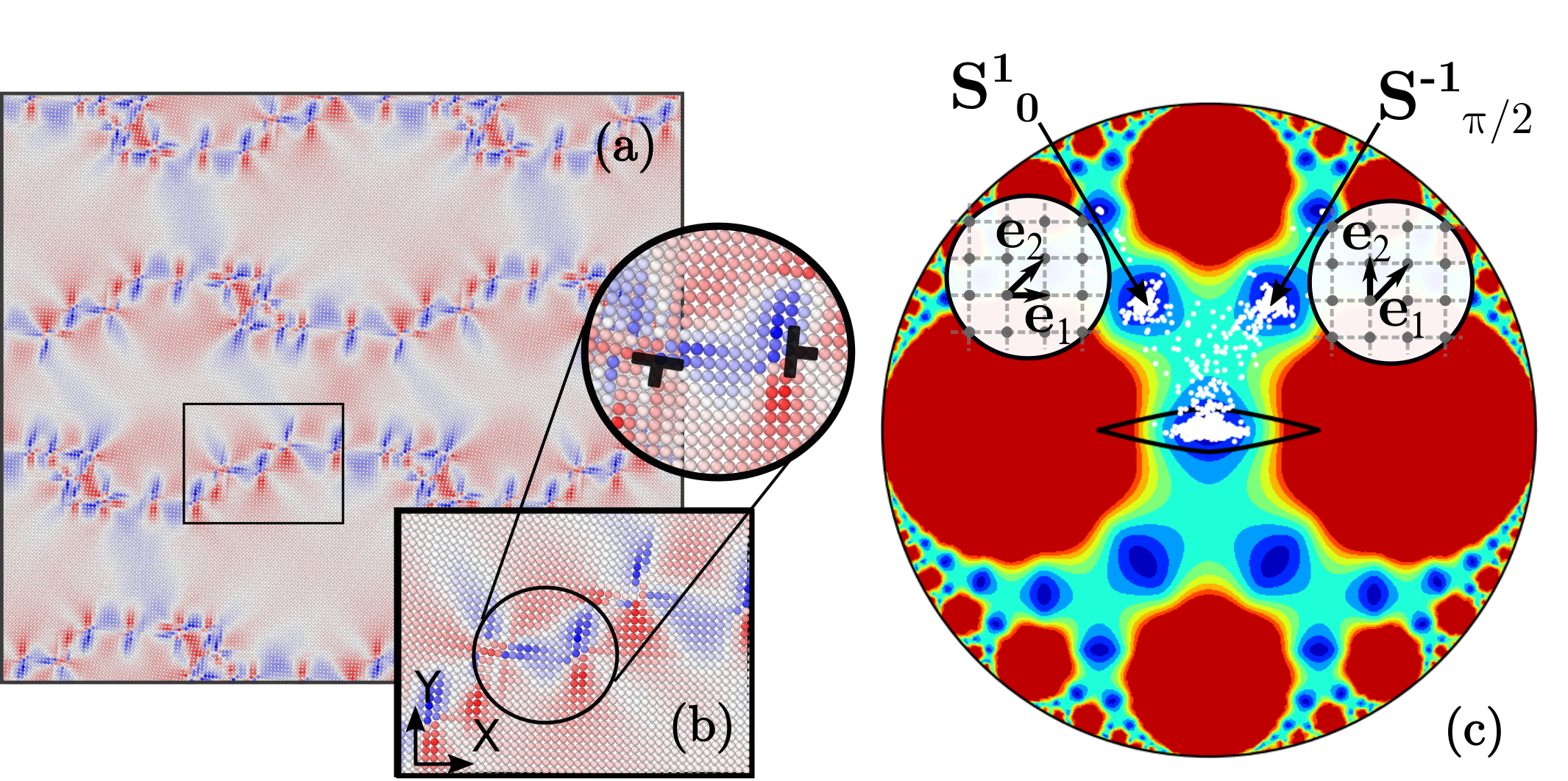}
\caption{\footnotesize{{\bf (a)} Post-instability pattern for the rhombic path ${\bf F}_{\diamond}$, colors  show the level of the  Cauchy stress $\sigma_{xy}$. {\bf (b)} Inset highlights the presence of both vertical and horizontal dislocations. {\bf (c)} Distribution of ${\bf C}_i$ points in configuration space show the dominant presence of the three wells $\bf S$, ${\bf S}^{1}_{0}$ and ${\bf S}_{\pi/2}^{-1}$. The elements on the low energy valleys connecting wells corresponds to  dislocation cores.} \label{fig:sq_soft_patt}}
\end{figure*}

\begin{figure*}[h!]
\centering
\includegraphics[width=.4 \textwidth]{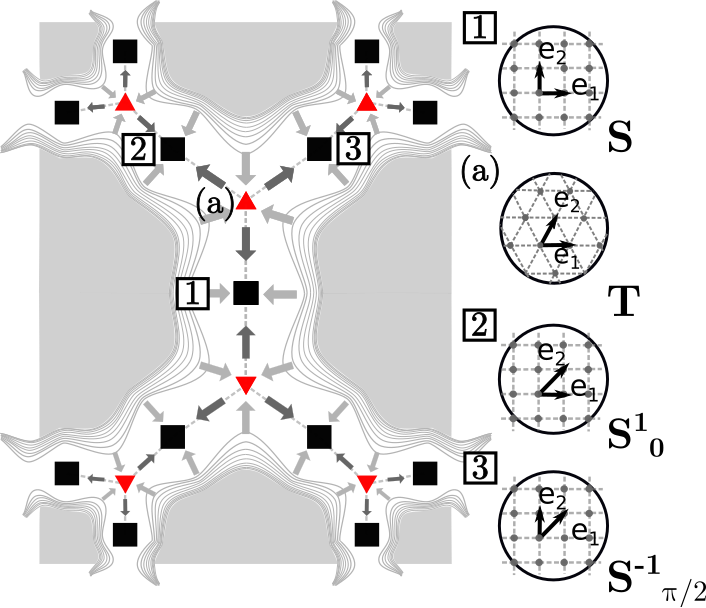}
\caption{\footnotesize{ Schematic representation of the saddle-like structure of the triangular phases in the square symmetry crystal. When the reference square configuration (1) is loaded along the energy valley, the systems encounters the triangular phase (a) and here splits along two different slip system, thus involving two additional wells (2) and (3). This splitting mechanism presents every time a square phase is loaded towards a triangular one.} } \label{fig:sq_soft_schem}
\end{figure*}

\paragraph{Square lattices.} We start with the case of a square lattice loaded along the 'soft' rhombic loading path ${\bf F}_\diamond$. The fragment of the post-instability pattern, shown in Fig. \ref{fig:sq_soft_patt}(a);  the colors in this image representing the physical space  indicate the level of the  Cauchy stress $\sigma_{xy}$. The observed   simultaneous activation of both   available slip systems is compatible with the   emergence  of  two unstable modes in the linear analysis which suggests   dislocation nucleation along the planes with two types of normals ${\bf n}_i$. We note the concurrent  initiation  of the horizontal and vertical slip systems has been  already observed in \cite{Salman2021-sn} for the case of the  simple shear loading path ${\bf F}_{\theta=0}$.This is not surprising since the  two paths corresponding to simple and pure shear   cross the stability  boundary in configurations which are very close to each other in $\bf C$ space.

The obtained dislocation pattern can be  understood further if we represent it in the  configurational space of metric tensors, see our in Fig. \ref{fig:sq_soft_patt}(b). In the homogeneous elastic state all configurational points were in the same location which depended parametrically on the loading parameter $\alpha$. After the effective yield surface was reached the configurational points spread over the configurational space with most of them concentrating in the three equivalent energy wells corresponding to  the reference  square lattice  ${\bf S}$, and the equivalent square lattices ${\bf F}_0(\alpha=1)$ and ${\bf F}_{\pi/2}(\alpha=-1)$  which differ from the reference lattice by  lattice invariant shears along the two perpendicular slip direction. Since the corresponding states have zero energy, such a localization indicates the formation of unloaded square lattice patches (grains)  which differ only by   rotation. The points outside the energy wells are mostly located inside the  energy valleys connecting the reference lattice  ${\bf S}$ with equivalent  configurations ${\bf F}_0(\alpha=1)$ and ${\bf F}_{\pi/2}(\alpha=-1)$, and corresponding to the horizontal and vertical dislocation core structures. Those structures are not exactly built as the pairs of  pure shear partials  studied above because they form  grain boundaries (dislocation walls) where dislocation interaction is strong.

We remark that the discussed   coupling between the slip systems is not postulated phenomenologically, as it is usually done in conventional continuum theories of crystal plasticity,  but emerges directly from the postulated global symmetry of the energy landscape. We illustrate this point in our Fig. \ref{fig:sq_soft_schem} where we show the zoom in on the schematic energy landscape in the configurational space around the reference energy well ${\bf S}$. This figure emphasizes the presence of the valleys which represent  the classical 'plastic mechanisms' and  direct the flow of configurational points away from the energetically expensive purely elastic deformation. It shows that an  exit from the narrow stability neighborhood of the point ${\bf S}$ (elastic domain) leads to the flow of the configurational points towards the degenerate saddle regions corresponding to the triangular lattice with the higher symmetry than the symmetry of the reference state; we note  the triangular lattice  automatically corresponds to a critical  point due to the global symmetry of the energy landscape.  

For instance, suppose that  the system is driven along the rhombic loading path ${\bf F}_\diamond$. It  is then forced directly towards the mountain pass around the point  ${\bf T}$ where the system is confronted with a (binary)  choice between moving either towards the (square)   well ${\bf F}_0(\alpha=1)$ or the (square)   well ${\bf F}_{\pi/2}(\alpha=-1)$ or, as in real numerical experiments, moving in both directions simultaneously while activating in this way both slip systems (here we are not talking about the configurational points that simply relax into the reference state).  In the case of a less symmetric loading path, like for instance, the simple shear path ${\bf F}_{\theta=0}$,  the choice will be  slightly  biased  with both slip systems still available due to the superior symmetry of the saddle region.  It is clear that when the system is loaded beyond the first avalanche, a succession of similar binary choice enhances the complexity in the developing pattern even further. 

%
%

  

\begin{figure*}[h!]
\centering
\includegraphics[width=.7\textwidth]{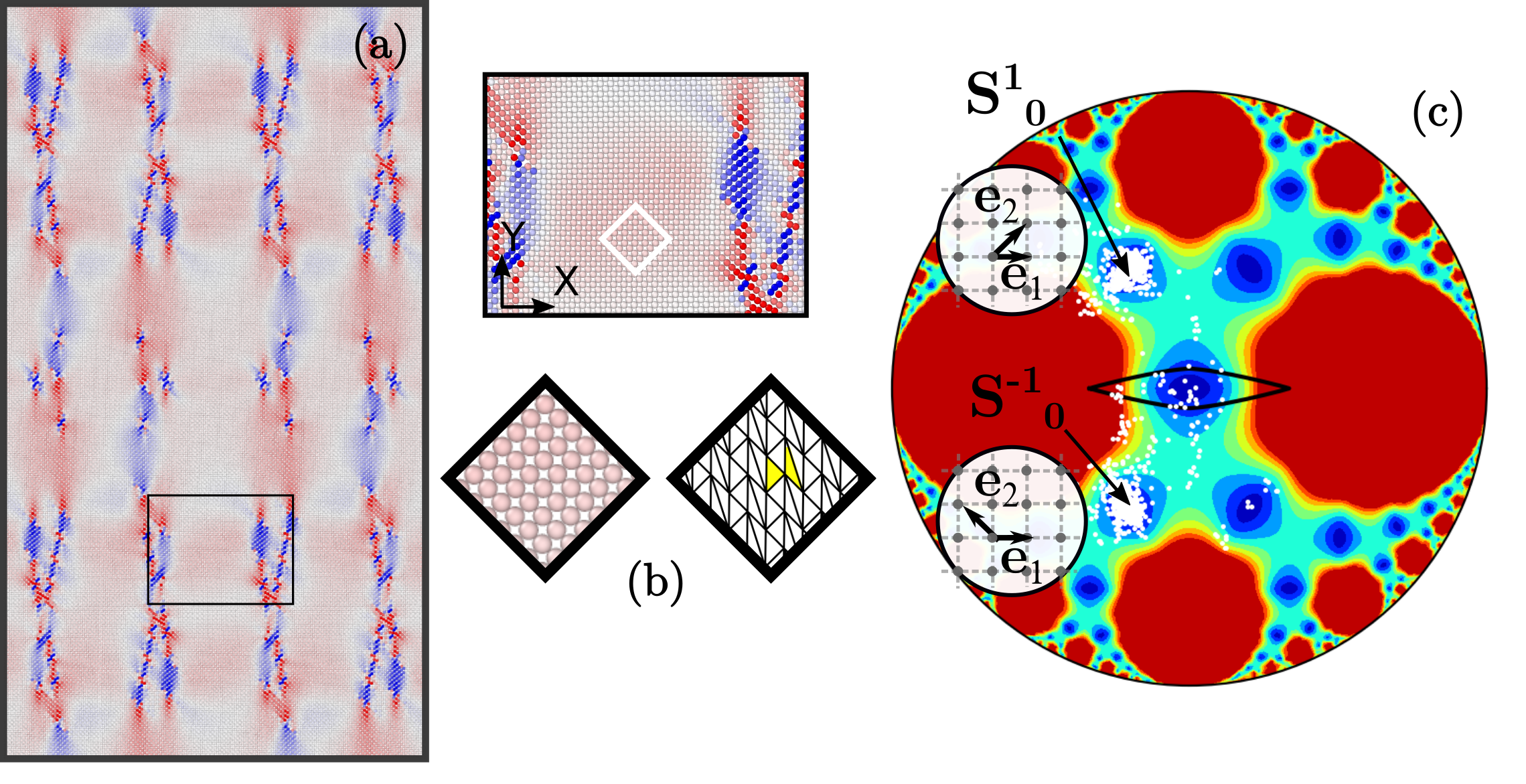}
\caption{\footnotesize{ {\bf (a)} Post-instability pattern for the rectangular path ${\bf F}_{\square}$, colors indicate the level of the Cauchy stress $\sigma_{xy}$. {\bf (a)}  The insets allow one to visualize the $\pi/4$ rotated structure,  the triangulation reveals the shearing mechanism behind such apparent rotation. {\bf (c)} Distribution of ${\bf C}_i$ points in configuration space show the splitting of the system between the wells ${\bf R}^{\pi/4}{\bf S}^{1}_{0}$ and ${\bf R}^{-\pi/4}{\bf S}^{-1}_{0}$.}\label{fig:sq_hard_patt}} 
\end{figure*}


We now discuss the  "hard" loading path   $\mathbf{F}_\square$ corresponding to driving through the imposed on the boundary affine  rectangular pure shear.  We recall that in this case  the square elements of the  reference lattice are  deformed elastically  into rectangles  with progressively higher energy cost. As we have also seen before,  the instability of the ensuing rectangular lattice leads to the formation of the  sheared layers oriented  perpendicular to the long axis of the rectangles which is a horizontal direction and with  their shear amplitude aligned with the vertical direction. The direction of the shear presents a binary choice between the (square) energy wells ${\bf S}_{1}^{0}={\bf F}(\theta=0,\alpha=1)$  and   ${\bf S}_{-1}^{0}={\bf F}(\theta=0,\alpha=-1)$ which suggests micro-twinning mechanism of instability. 

\begin{figure}[h!]
\centering
\includegraphics[width=.75\columnwidth]{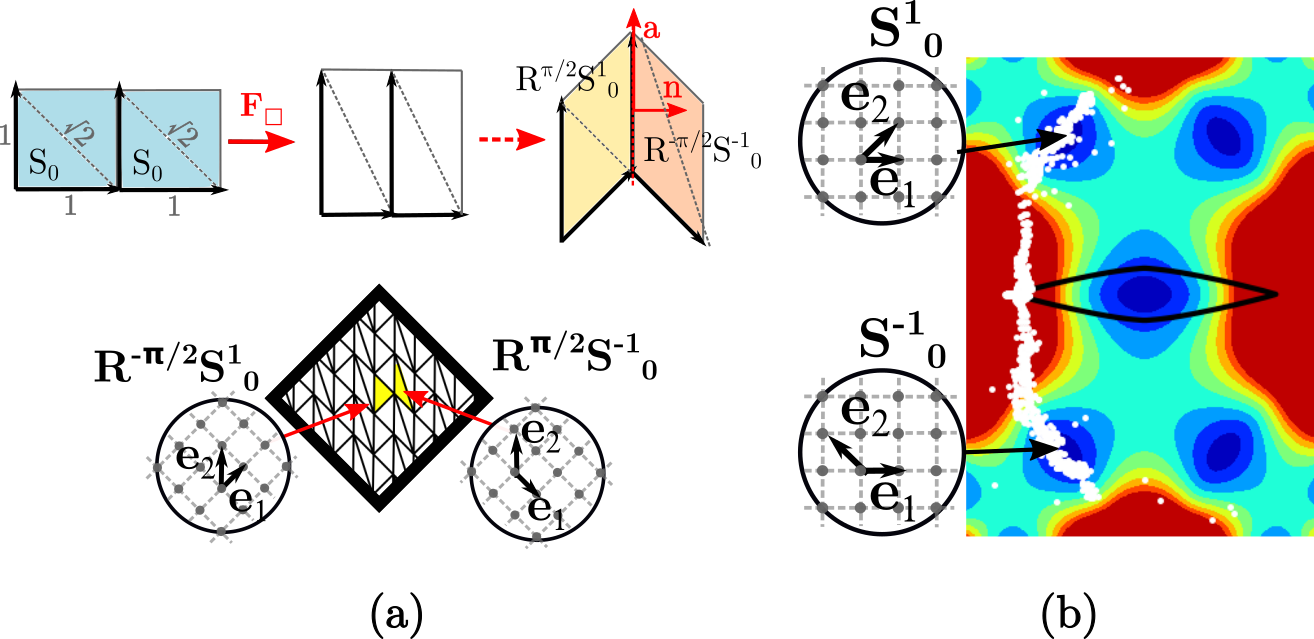}
\caption{\footnotesize{{\bf (a)} The twinning mechanism behind the apparent rotation. The instability   develops with a   redistribution of the elements between the energy wells ${\bf R}^{\pi/4}{\bf S}^{1}_{0}$ and ${\bf R}^{-\pi/4}{\bf S}^{-1}_{0}$. {\bf (b)} A snapshot of the developing instability, showing the early evolution of the system towards the two equivalent wells.}} \label{fig:sq_hard_patt_bis}
\end{figure}

The post avalanche configuration obtained  in the corresponding numerical experiment is illustrated  in  Fig.~\ref{fig:sq_hard_patt}. The analysis of the physical state 
 reveals the system size  pattern 
  where patches of the  original square lattice structure appear  to be  rotated at $\pi/4$. The  dislocation rich   high energy defects    serve again as   the  boundaries separating these  grains, see Fig. \ref{fig:sq_hard_patt}(a). The deformed configuration of the elements inside the grains  shows that the apparent rotation is produced by   the fine lamination of the (almost) unloaded states  from the  different energy  wells ${\bf S}_{1}^{0}$ and ${\bf S}_{-1}^{0}$, see   Fig. \ref{fig:sq_hard_patt}(b). The implied   two-well redistribution  is clearly visible in the configurational space, shown in Fig. \ref{fig:sq_hard_patt}(c),  where we see that these two wells are almost equally populated with almost no elements flipping back into the original energy well ${\bf S}$.
This type of accommodation through inelastic rotation can be easily understood if we observe that the two sheared state configurations constituting the micro-twin laminate, ${\bf F}=\mathbf{R}(\pi/4){\bf S}^{0}_1 $ and ${\bf G}=\mathbf{R}(-\pi/4){\bf S}^{0}_{-1}$ satisfy the compatibility condition    \cite{Pitteri2002-rm} 
$
\mathbf{F}={\bf I}+\left(\mathbf{a}\otimes\mathbf{n}\right){\bf G},
$
where   ${\bf a}^T=\left(0,2\right)$ and ${\bf n}^T=\left(1,0\right)$, see Fig. \ref{fig:sq_hard_patt_bis}(a).   Note that the normal to  the twinning plane  $\bf n$ coincides with the instability direction predicted by our approximate  stability analysis.

In Fig. \ref{fig:sq_hard_patt_bis}(b) we show the distribution of the configurational points immediately following the onset of instability, when the avalanche is only unfolding. It suggests that a highly inhomogeneous  configuration precedes the development of the micro-laminates disguised as uniformly rotated grains. The eventual equilibration is achieved through the advancement of a dynamic front.  Inside such a  transition front  the apparent   rotation of the lattice is achieved through transverse motion of dislocations which nucleate inside the computational domain but ultimately annihilate  on the boundary \cite{baggio-arxiv}.

\begin{figure*} [h!]
\centering
\includegraphics[width=.7\textwidth]{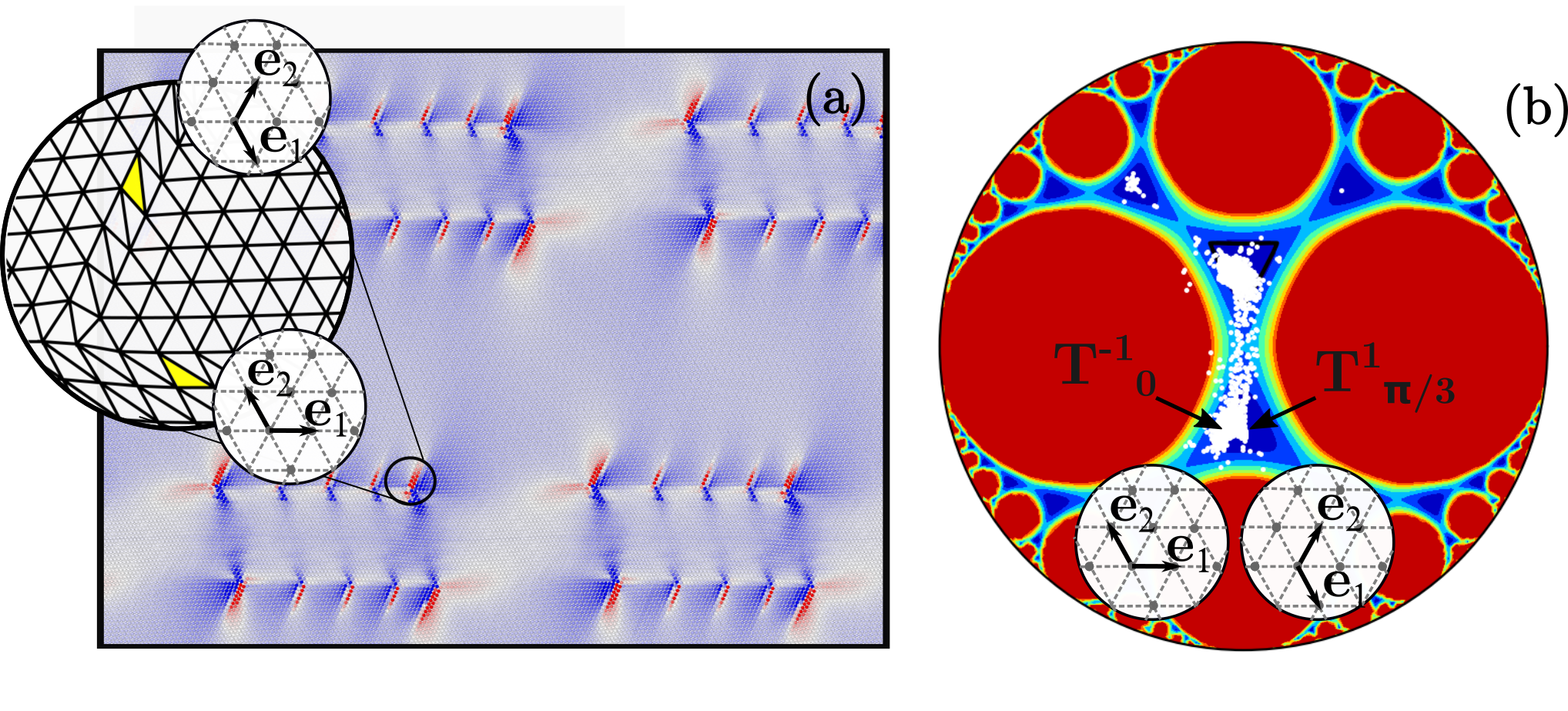}
\caption{\footnotesize{{\bf (a)}   Post-instability pattern observed on the rhombic path $\bar{\bf F}_{\diamond}$, colors indicate the level of the  Cauchy stress $\sigma_{xy}$. The inset shows a detail of the triangulation revealing the activation of two slip systems. {\bf (b)} Distribution of ${\bf C}_i$ points in configuration space show that the majority of points lies in the low energy valleys connecting $\bf T$ with ${\bf T}^{-1}_{0}$ and ${\bf T}_{\pi/3}^{-1}$. Since these two configurations differs by a rigid rotation only, the paths are overlapping. }} \label{fig:hex_soft_patt}
\end{figure*}

\paragraph{Triangular lattices.} Consider now  the 'soft' pure shear loading protocol $\bar{\bf F}_{\diamond}$ applied to a  triangular lattice. In Fig. \ref{fig:hex_soft_patt}(a)  we show a fragment of the  post avalanche pattern in the physical space; the corresponding distribution of the configurational points is presented in Fig. \ref{fig:hex_soft_patt}(b). As in the case of 'hard' pure shear loading of a square crystal, here we again see the emergence of slip  on  two slip systems (out of three available in general).   We recall that also according to the linear stability analysis two slip directions are activated simultaneously.  Interestingly, and differently from the case of  square symmetry, in our numerical experiments involving triangular lattices loaded by simple shears along the closest crystallographic directions to $\bar{\bf F}_{\diamond}$, for instance $\bar{\bf F}_{\pi/3}$ or $\bar{\bf F}_{0}$, such double activation of two slip systems does not take place \cite{ Salman2021-sn}.  This is related to a  structurally different organization of the low energy valleys around the reference states for  square and triangular crystals  and the resulting different mismatch   between the critical stability thresholds  along simple and pure shear loading paths.

Note first that in the case of triangular lattices, the loading paths $\bar{\bf F}_{\pi/3}$ and $\bar{\bf F}_\diamond$ intersect the boundary of the elastic (stability) region in the configurational points that are rather  distant from the point where such crossing takes place for  the  pure shear path  $\bar{\bf F}_{\diamond}$  while   in the case of  square lattices  all three paths cross the stability boundary at almost the same point (compare  Fig. \ref{fig:en_hex_saddle} and  Fig. \ref{fig:en_sq_saddle}). In other words, the triangular lattice, driven along the path $\bar{\bf F}_\diamond$, becomes unstable in the middle of the energy valley, quite late vis a vis  the instability under  the simple shear protocols $\bar{\bf F}_{0}$ and  $\bar{\bf F}_{\pi/3}$.  The fact that this happens close to the saddle {\bf S} facilitates the coupling between the slip systems oriented at the angles $\theta  = \pi/3$ and $\theta= 0$.


Note next that while the simple shear loading  paths $\bar{\bf F}_{\pi/3}(\alpha)$ and $\bar{\bf F}_{0}(-\alpha)$ are distinct,   they intersect not only at $\alpha=0$ (at the  reference energy well ${\bf T}$) but also at $\alpha=\gamma^2$ ( at the equivalent energy well ${\bf T}^{-1}_{2\pi/3}$,  where the pure shear   loading path $\bar{\bf F}_\diamond$ ultimately leads. The fact that the two simple shear paths are ultimately getting closer to the main driving direction  contributes to the ultimate activation of both slip systems.

 
\begin{figure}[h!]
     \centering
     \begin{subfigure}{0.65\textwidth}
         \centering
         \includegraphics[width=\textwidth]{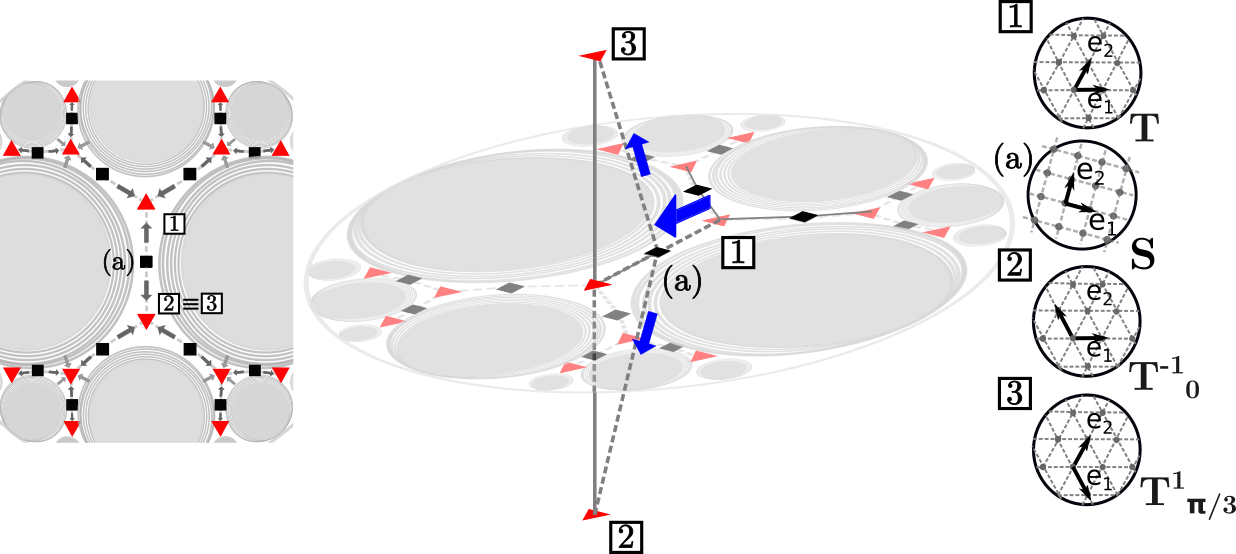}
         \caption{}
         \label{fig:hex_sadd_ssch}
     \end{subfigure}
\hspace{0.01\textwidth}
     \begin{subfigure}{0.3\textwidth}
         \centering
         \includegraphics[width=\textwidth]{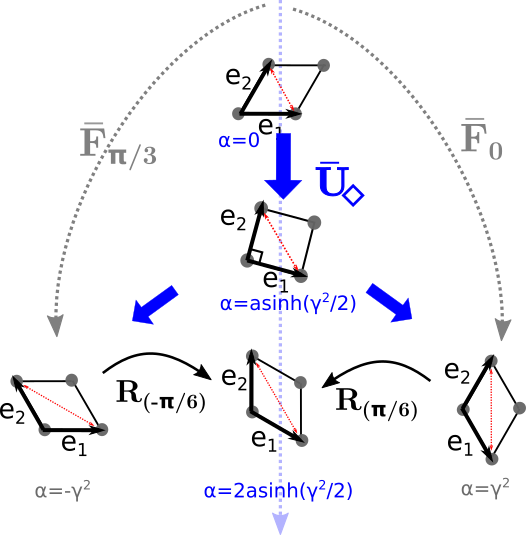}
         \caption{}
        \label{fig:dislo_patt_sch}
     \end{subfigure}
\caption{\footnotesize{{\bf (a)} Schematic representation of the saddle-like structure of the square phases in the triangular symmetry crystal. An extra dimension (here showed with vertical lines) needs to be included to observe rotated wells along the same orbit, that in these case intervene when considering the shears aligned with the crystallographic planes (oriented). Here we consider the reference triangular phase (1), loaded towards the square phase (a). The system ends up activating two slip system whose corresponding wells are distinguished by a rigid rotation. {\bf (b)} Pure shear (Eq. \ref{eq:pure_shear_fat}) and simple shears $\bar{\bf F}_{\pi/3}(\alpha=\gamma^2)$, $\bar{\bf F}_{0}(\alpha=-\gamma^2)$ leads to formation of  different patterns comprised of  energy wells distinguished solely by a rigid rotation}; {\bf (c)}  The  symmetric rotations  disguising  two  simple shears as  one  pure shear  \label{fig:hex_saddle_sch}
}
\end{figure}

We remark that, while  activation of the two slip systems is   clearly visible in physical space, see Fig. \ref{fig:hex_soft_patt}(a),  it is less apparent  from the spreading  of the cloud of configurational points in the space of metric tensors, see  Fig. \ref{fig:hex_soft_patt}(b) 
 where we see that at the saddle ${\bf S}$ about half of the  elements flip back to the original well ${\bf S}$ while another half advances to the new well ${\bf T}^{-1}_{2\pi/3}$.  
 However,   the two states  $\bar{\bf F}_{\pi/3}(\alpha=\gamma^2)$ and $\bar{\bf F}_{0}(\alpha=-\gamma^2)$, which  occupy the same point ${\bf T}^{-1}_{2\pi/3}$ in our conventional configurational space of metric tensors, differ by a rigid rotation. 

\begin{figure*} [h!]
\centering
\includegraphics[width=.8\textwidth]{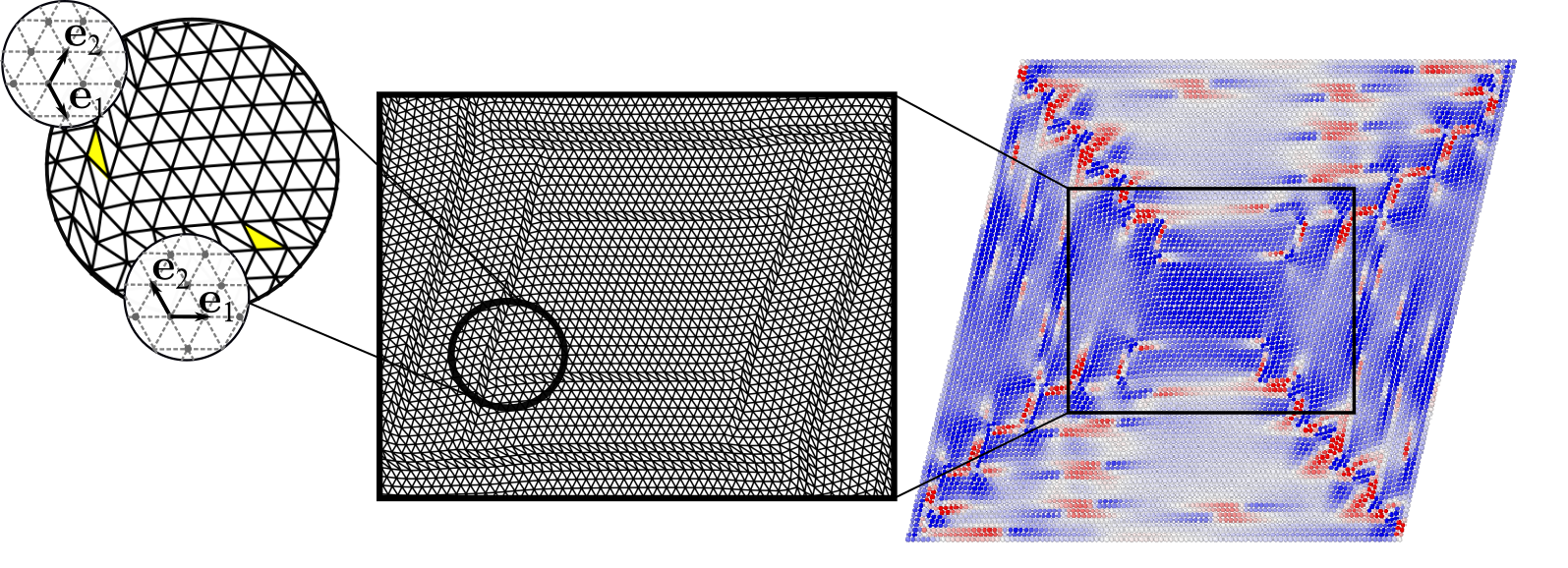}
\caption{\footnotesize{Post-instability pattern for the rhombic path $\bar{\bf F}_{\diamond}$, observed with fixed boundary conditions.}} \label{fig:hex_soft_patt_fix}
\end{figure*}

To explain this point we recall that  even though the two configurations may belong to the same energy well, they may correspond to different points of the orbit of this well and formed by rotations which leave the metric tensor unchanged~\cite{Pitteri2002-rm}. 
In Fig. \ref{fig:hex_saddle_sch} we illustrate by a scheme, a likely mechanism of the simultaneous activation of the two slip systems. While the horizontal plane in this scheme represents our  conventional configurational space of metric tensors   (see also the inset to the left of the scheme), the vertical direction mimics  a  one-parameter space of rigid rotations which we neglected in all previous considerations.  When the triangular lattice  ${\bf T}$, marked as  (1),  evolving  along the energy valley,  reaches the saddle describing the square lattice ${\bf S}$, marked as  (a),  two rotations of the same amplitude but of  different signs start to develop as the system continues to evolve along the energy valley down from the saddle ${\bf S}$ towards the energy well ${\bf T}^{-1}_{2\pi/3}$,  while  fully maturing as symmetric slips along the  close packed directions  $\theta=\pi/3$ and $\theta=0$.

In Fig. \ref{fig:hex_soft_patt_fix} we show, for comparison, the  post-avalanche pattern in the same setting but with fixed affine boundary conditions. While it has the same elementary local dislocational patterns as in the case  of the periodic  boundary conditions,   the global organization is largely  shaped by the influence of the boundaries. This and other finite size effects will be considered in more detail in a separate publication.






%
%
%
%

 %
\begin{figure*} [h!]
\centering
\includegraphics[width=.6\textwidth]{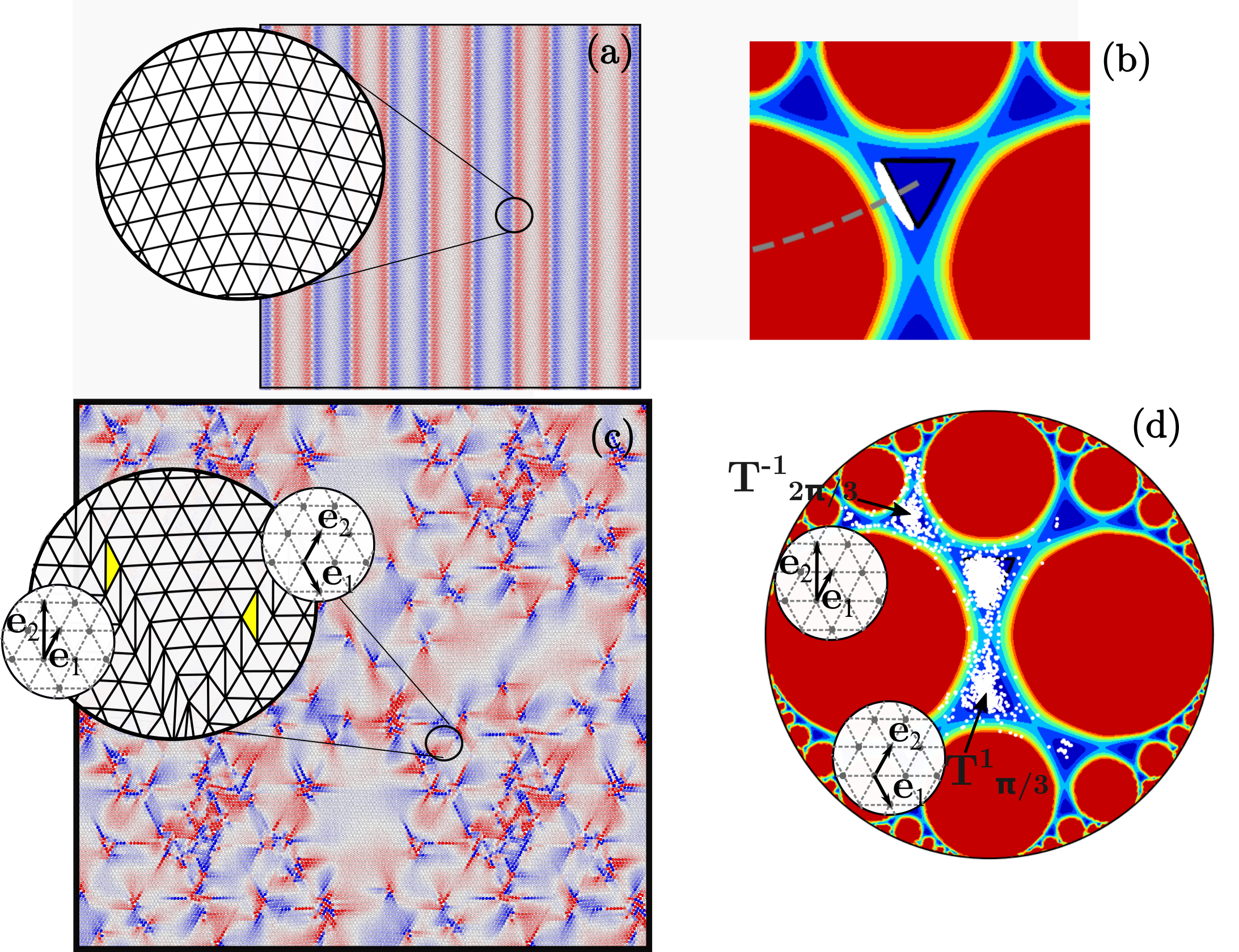}
\caption{\footnotesize{{\bf (a)} The emerging inhomogeneous  configuration  at the point of  instability, characterized by a pattern  of  weakly  rotated triangular lattices; the corresponding distribution in configuration space is illustrated in {(\bf b)}. Below: the post-avalanche structure takes the form of a double dislocation nucleation along crystallographic planes $\pi/3$ and $2 \pi/3$. In {\bf (c)} we show the pattern in the physical space, along with an inset of the triangulation, while in {\bf (d)} we show the distribution of ${\bf C}_i$ among finite elements in configuration space. }} \label{fig:hex_hard_patt}
\end{figure*}
%

 
Finally, consider a  triangular lattice driven using the 'hard' loading protocol  $\bar{\bf F}_\lozenge$  representing  rhombic pure shear. In this case the system is moved   away from the energy well  ${\bf T}$ along   the steep energy  hill  acquiring progressively increasing  elastic energy. As in the case of  the 'hard' pure shear loading of a square lattice, the eventual instability of the homogeneous configuration of the elastically deformed  triangular lattice   can be expected to resolve into a symmetric (micro-twin?) mixture of the two   triangular lattices corresponding to the energy wells ${\bf T}^{-1}_{2\pi/3}$ and ${\bf T}^{1}_{\pi/3}$.

 The results of our numerical experiment are reported in Fig. \ref{fig:hex_hard_patt}. We first show in Fig. \ref{fig:hex_hard_patt}(a-b) the initial (elastic) stage of the instability when the system still remains in the vicinity of the reference state  ${\bf T}$ while developing periodic modulation oriented in accordance with the  predictions produced by the theoretical study of the linear elastic instability. While such   modulation does not involve  the anticipated activation of the two symmetry related  energy wells,  the   increasingly pronounced periodic patterning   resembles a somewhat blurred micro-twin structure involving  a  mixture of the energy wells ${\bf T}^{-1}_{2\pi/3}$ and ${\bf T}^{1}_{\pi/3}$ , see Fig. \ref{fig:hex_hard_patt}(a). These two wells are in fact compatible and can in principle  mix (laminate) to produce a rotation of the original triangular lattice. The corresponding twinning equation is  analyzed in Appendix \ref{app:twinning}.

However,  under further loading, this highly ordered  inhomogeneous configuration does not evolve into an organized micro-twin laminate as in the case of  similar loading protocol for square crystals.  Instead, at the advanced stage of the avalanche,  some elements flip back to the reference energy well ${\bf T}$ and the incipient periodic  pattern breaks down  with a massive nucleation of dislocations  of the two types:  connecting either the wells   ${\bf T}$ and ${\bf T}^{-1}_{2\pi/3}$ or the wells  ${\bf T}$ and ${\bf T}^{1}_{\pi/3}$.  During this breakdown process   we observe sharp  drop in stress and energy as the two slip systems are activated simultaneously. The avalanche ends with a  formation of    a complex arrangement of self-locked dislocations, see Figure \ref{fig:hex_hard_patt}(c). The final configuration in the space of metric tensors is represented by the three symmetric energy wells almost equally populated. 
\begin{figure*} [h!]
\centering
\includegraphics[width=.8\textwidth]{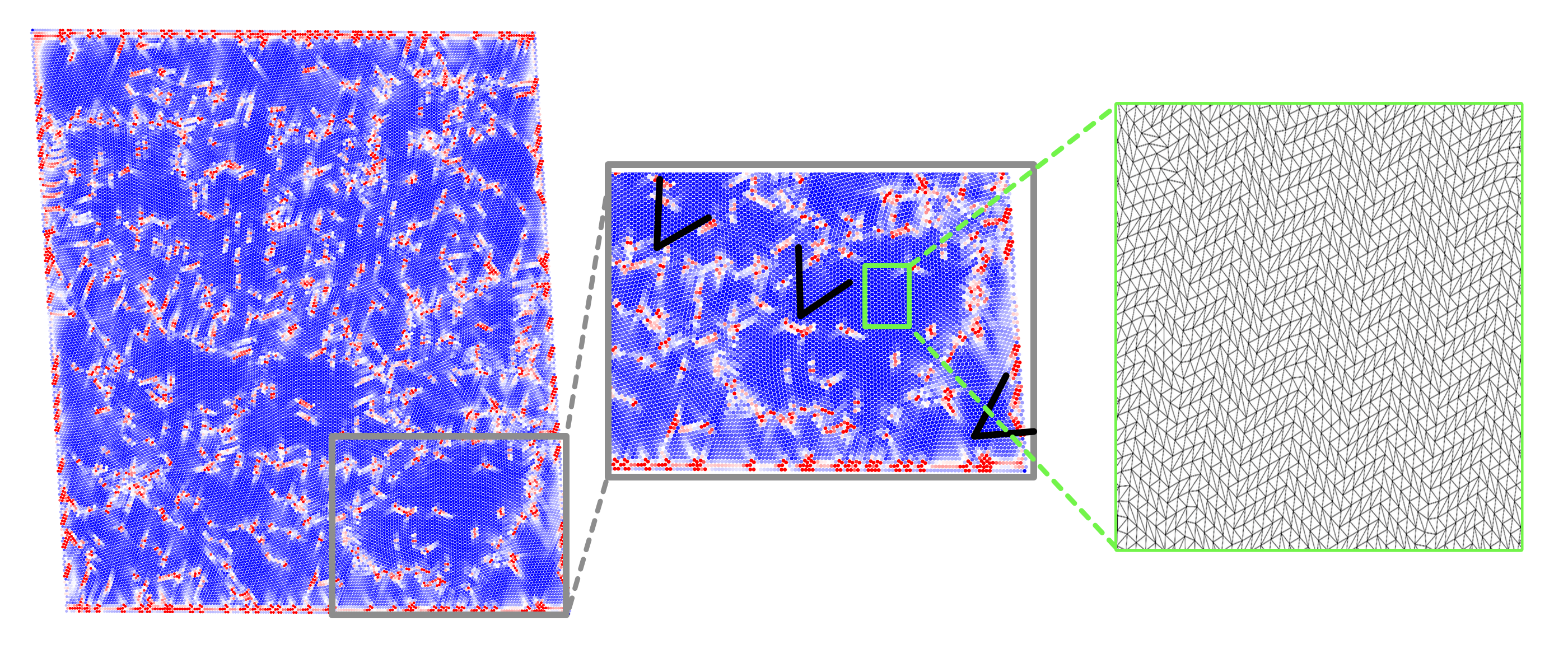}
\caption{\footnotesize{Post-avalanche  pattern for a  triangular lattice deformed along the simple shear loading path  with  $\theta=54$ degrees, see Eq. \ref{path:tri}, which shows  variously oriented  dislocation-free grains }} \label{fig:app_fig}
\end{figure*}

The observed differences in the character of the collective dislocation nucleation phenomenon  along the 'hard' loading paths  in   triangular and square lattices are  probably related to the higher symmetry of the former. Thus, in triangular lattices due to the more 'compact' structure of the effective yield surface, the instability of a homogeneous states takes place at lower levels of elastic energy  which is then less available for the restructuring of the lattice. Therefore, instead of micro-twinning, aimed at the reduction of  the energy globally,  the system mimizes the energy locally by producing an intricate network of self-jammed dislocational entanglements. 
In other words the  breakdown of metastability  simply does not release  enough energy to access the micro-twinned configuration, which requires major rearrangement.  

Interestingly, our numerical experiments showed that using a specially designed  loading protocol,  the local micro-twinning can be achieved, see Fig. \ref{fig:app_fig}. Here one can see that the overall pseudo-rigid rotation inside a grain can be reached by complex micro-twinning which involves coexistence of the  three unloaded triangular lattices corresponding to the bottoms of the energy wells, $ {\bf T}$,  ${\bf T}^{-1}_{2\pi/3}$  and ${\bf T}^{1}_{\pi/3}$,  which are separated by   semi-coherent grain boundaries oriented at either zero or 60 degrees, in accordance with the theoretical    prediction  made in our   Appendix \ref{app:twinning}.


%

\section{Conclusions}
\label{sec:conc}

In this paper we have presented new insights on homogeneous nucleation of dislocations  in 2D  pure crystals by emphasizing the collective nature of this phenomenon. These insights became possible due to the use of the novel  mesoscopic tensorial model (MTM) of crystal plasticity which combines the advantages of pseudo-macroscopic description of plastic flows in terms of stresses and strains with the ability to describe short range interaction of dislocations and even resolve the crystallographic symmetry sensitive aspects of the structure of their cores. In contrast to some other  mesoscopic approaches, the MTM does not require any dislocation-specific phenomenological entries and relies almost exclusively on the global symmetry of the lattice. This symmetry goes beyond the conventional point group and accounts in a geometrically exact way of lattice invariant shears.

The phenomenon of  the homogeneous nucleation of dislocations presents a convenient  background for testing the access of  MTM to the crucial mesoscopic features of crystal plasticity. Previously, such nucleation in 2D was modeled  as a localized  event  resulting in  the  formation of a  topologically neutral pair of dislocations of opposite signs.  Here we show that  in the absence of defects and inhomogenities, the dislocation nucleation in pristine simple crystals  unfolds as a system size avalanche. Due to the dominance of long range elastic interactions,  it emerges as a collective phenomenon, involving a large number of dislocations,  and leading to the formations of intricate  patterns of global nature. We  showed that some  important  peculiarities    of such patterns may sensitively depend on the crystallographic  symmetry of the lattice.

To highlight the importance of crystal symmetry in the process of homogeneous nucleation of dislocations we considered two main classes of simple lattices amenable to modeling in 2D: the lower symmetry  square lattice and the higher symmetry triangular lattice. The possibility of defining general loading protocols allowed us to compare for both types of lattices the two archetypal loading paths: along the maximally 'soft' direction and along the maximally 'hard' direction.  

Driving in the  'soft' direction reveals  a non-trivial coupling between several  slip  systems allowing the crystal to accommodate the applied loading by forming a relatively regular patterns of dislocation walls.  The important role in such coupling is played by the 
metastable phases: triangular lattice  ${\bf T}$ during the plasticity of square crystal   $\bf S$ and vice versa.   While in the case of plasticity of square crystal the implied branching of the energy valleys at the location of the triangular lattice  $\bf T$ is immediately apparent, the situation is less simple in the case of plasticity of  triangular crystals where the branching at the location of the square lattice  $\bf S$  is  between the different  points of the  orbit of the same lattice  ${\bf T}^{1}_{\pi/3}\cong {\bf T}^{-1}_{0}$. 

Instead, driving in a 'hard'  direction,   produces  in crystals with lower symmetry a regular pattern of  mutually misoriented patches  (or grains) where plastic deformation takes the form of micro-twinning disguised as rigid rotation \cite{baggio-arxiv}. Thus the  collective  nucleation  of dislocation in the case of square crystals,    ultimately resulting in   a  formation of   laminates,   proceeds through  the propagation of a front. The latter    involves the transverse motion of individual dislocations which are finally   expelled to the boundary of the crystal  leaving behind a  fully unloaded but inelastically rotated  original lattice. Such perfectly organized pattern  fails to develop  in  triangular crystal, where  it  is replaced by a more complex network of jammed dislocation self-locks. Apparently, due to  the higher symmetry of the crystal in this case,  the dislocation generating instability takes place at the lower levels of stress  which prevents global rearrangement  replacing it with more local self-organization of  individual slips. 

Our exploratory study shows the  strength of  the MTM  in  dealing with the micro-structural aspects of  crystal plasticity. This model can be potentially developed with no phenomenology  at all if the periodic potential is constructed by ab initio methods. The natural future target of the model is the study of the mechanical  fluctuations accompanying plastic yield. To be realistic the model should be  moved from 2D to  3D where it should be able to reproduce the experimentally observed   peculiarities of plastic fluctuations  in FCC, BCC and HCP crystals.

 

\section{Acknowledgments}
The authors acknowledge helpful discussions with G. Zanzotto  and C. Conti at the stage of  the development of the MTM theory. O. U. S. is supported partially by the grant ANR-18-CE42-0017-03, O. U. S. and R.B. were supported by the grants ANR-19-CE08-0010-01, ANR-20-CE91- 0010, and L. T. by the grant ANR-10-IDEX-0001-02 PSL.

\begin{appendices}

\section{Appendix: Numerical algorithm}
 \label{alg:hudson}
 \begin{algorithmic} [1] 
   \scriptsize
\State Generation of finite element mesh  of the body $ \Omega$ and identification of boundary nodes on the boundary of the body $\partial \Omega$.
\State Initialization of the displacement vector $\bf u=0$  for all nodes $a$.
\State Set  loading through the displacement ${\bf u}(\alpha)= (\bar{\bf F}(\alpha)-\mathbb{I}){\bf x}$ for all nodes $a$ on the boundary of the body $\partial \Omega$, where $\bar{\bf F}(\alpha)$ is the applied deformation gradient with amplitude $\alpha$. 
\State{\textit{Start the iterative L-FBGS minimization algorithm:} }
\State{ Construct a deformation gradient ${\bf F}$ in each finite element.}
\State{   Construct a metric tensor  ${\bf C} = {\bf F}^T {\bf F}$  in each finite element.}
\State{  Perform Lagrange reduction to calculate the reduced metric tensor  ${\bf C}_D$ and the ${\bf m}$ matrix  in each finite element.}
\State{ Calculate the first     Piola-Kirchhoff stress tensor.}
\State{ Obtain nodal forces.}
\State{ Obtain the total strain energy.}
\State{ Obtain the new displacement vector $\bf u^t$ at iteration $t$ such that  $W^{t}<W^{t-1}$ }

\State{   Ends  minimization  at iteration $t$ when $W^{t}-W^{t-1}<tol$ }
\State{{\textit{Start Newton algorithm  with the displacement vector $\bf u^t$ obtained after the termination of L-FBGS minimisation algorithm} }  }
\State{ Construct a deformation gradient ${\bf F}$ in each finite element.}
\State{ Construct a metric tensor  ${\bf C} = {\bf F}^T {\bf F}$  in each finite element.}
\State{ Perform Lagrange reduction to calculate the reduced metric tensor  ${\bf C}_D$ and the ${\bf m}$ matrix  in each finite element.}
\State{ Calculate the tensor  $\bf A$.}
\State{ Calculate the stiffness matrix  $\bf K$  and the residual forces  ${\bf R}$.}
\State{ Perform a Newton step.}
\State{ Obtain the new displacement vector $\bf u^t$ at iteration $t$ such that  the vector norm of residual forces $|{\bf R}^{t}|<|{\bf R}^{t-1}|$.}

\State{ Ends  the Newton-Raphson   at iteration $t$ when $|{\bf R}^{t}|-|{\bf R}^{t-1}|<tol$ }
\State{\textit{Increase the loading amplitude:} $\alpha\rightarrow \alpha+\delta\alpha$}
\State{Go to step 3}

 \end{algorithmic}
\section{Appendix: Twinning relations}
 \label{app:twinning}


Suppose that  the constant deformation gradients $\bf G$ and $\bf H$ correspond to two equivalent  minima of  the strain-energy $\phi(\bf C)$. To generate piece wise affine continuous deformation, across  an  invariant discontinuity plane they must  satisfy on such a plane  the  kinematic (Hadamard) compatibility conditions \cite{Pitteri2002-rm}:
\begin{equation}
\mathbf{R}\mathbf{H}=\mathbf{G}+\mathbf{a}\otimes\mathbf{n}^*=\mathbf{G}\left(\mathbf{I}+\mathbf{a}^*\otimes\mathbf{n}^*\right) =\left(\mathbf{I}+\mathbf{a}\otimes\mathbf{n}\right){\bf G}
\label{eq:twin_compatibility}
\end{equation}
where ${\bf R}\in SO(2)$ is a rotation. The Eulerian vector $\bf a$ (normal to the discontinuity plane) and covector $\bf n$ must satisfy  ${\bf a}\cdot{\bf n}=0$; their Lagrangian counterparts are  $\mathbf{a}^*=\mathbf{G}^{-1}\mathbf{a}$ and  $\mathbf{n}^*=\mathbf{G}^T{\mathbf{n}}$. 
If we assume further  that $\det \mathbf{H}=\det \mathbf{G}=1$ and exclude  reflections,  the  deformation gradients satisfying (\ref{eq:twin_compatibility}) form  a  mechanical twin.  If, in addition,  the rotation $\mathbf{R}$ belongs to the point group of the lattice, such twinning structure  produces the  undistorted zero energy configuration.  The resulting microtwinned laminates  are  sometimes referred to as \emph{pseudotwins} \cite{Pitteri2002-rm}.

The twinning equation \eqref{eq:twin_compatibility} was studied extensively, see for instance \cite{Forclaz1999-hu}. It was shown   that (\ref{eq:twin_compatibility}) admits either no solutions  or two solutions.  
The two solutions exist when the matrix  $\mathbf{G}^{-T}\mathbf{H}^T\mathbf{H}\mathbf{G}^{-1}\neq{\bf I}$ and its ordered eigenvalues $\mu_1<\mu_2$ are such that $\mu_1\mu_2=1$. In that case, the two solutions are given explicitly by the formulas:
\begin{align}
\mathbf{a}&=\rho\left(\sqrt{\frac{\mu_2(1-\mu_1)}{\mu_2-\mu_1}}\mathbf{v}_1+\kappa\sqrt{\frac{\mu_1(\mu_2-1)}{\mu_2-\mu_1}}\mathbf{v}_2 \right)\;,\\
\mathbf{n}&=\frac{1}{\rho}\left(\frac{\sqrt{\mu_2}-\sqrt{\mu_1}}{\sqrt{\mu_2-\mu_1}}\right)\left(-\sqrt{1-\mu_1}\mathbf{v}_1 +\kappa\sqrt{\mu_2-1}\mathbf{v}_2 \right)\;,
\end{align}
where $\mathbf{\hat v}_1$ and $\mathbf{\hat v}_2$ are the normalized eigenvectors of $\mathbf{G}^{-T}\mathbf{H}^T\mathbf{H}\mathbf{G}^{-1}$, $\rho>0$ is a constant ensuring that $|{\bf n}|=1$ and $\kappa=\pm 1$. Once  $\bf a$ and $\bf n$ are known, the rotation $\bf R$ can be obtained directly from \eqref{eq:twin_compatibility}. 

\noindent First, we  consider the compatibility of the two nearest wells reachable  by deforming  the original triangular phase  using the deformation gradients:
\begin{mycases}
\item  ${\bf H}=  \begin{pmatrix}
1 & \gamma^2\\
0 & 1 
\end{pmatrix} $ and  ${\bf G}= \begin{pmatrix}
1 & -\gamma^2\\
0 & 1 
\end{pmatrix}.$ They correspond  to the zero degree shear defined in Eq. \ref{eq:shear}  such that  ${\bf H} = {\bf F}(\gamma^2,0)$ and ${\bf G} = {\bf F}(-\gamma^2,0)$. 

\item  ${\bf H}=  \begin{pmatrix}
1 & \gamma^2\\
0 & 1 
\end{pmatrix} $ and  ${\bf G} =  \begin{pmatrix}
0.5 & \sqrt{3}/6\\
-\sqrt{3}/2 & 1.5 
\end{pmatrix} $. 
The phase ${\bf G}$   is accesible by deforming the original triangular phase by ${\bf G} = {\bf F}(-\gamma^2,\pi/3)$. 
\item  ${\bf H}=  \begin{pmatrix}
0.5 & \sqrt{3}/6\\
-\sqrt{3}/2 & 1.5 
\end{pmatrix} $ and  ${\bf G} =  \begin{pmatrix}
0.5 & -\sqrt{3}/6\\
\sqrt{3}/2 & 1.5
\end{pmatrix} $. 
The phase ${\bf G}$   is accesible by deforming the original triangular phase by ${\bf G} = {\bf F}(-\gamma^2,2\pi/3)$. 
\item   ${\bf H}=  \begin{pmatrix}
1 & -\gamma^2\\
0 & 1 
\end{pmatrix} $ and  ${\bf G} =  \begin{pmatrix}
0.5 & -\sqrt{3}/6\\
\sqrt{3}/2 & 1.5 
\end{pmatrix} $.

\end{mycases}
	
We found that for each of the cases described above, the twinning equation admits solutions  summarized below for each case:
\begin{mycases}
\item   Solution corresponding to  $\kappa=1$  is given by 
  \begin{equation}
\mathbf{a}^T=\lbrace -1.74574, 1.51186 \rbrace \quad \mathbf{n}^T=\lbrace 0.654654, 0.755929 \rbrace,
\end{equation} 
and the corresponding rotation angle is 98.2132 degrees. . For $\kappa=-1$, the solution is different
  \begin{equation}
\mathbf{a}^T=\lbrace -2.3094,0 \rbrace \quad \mathbf{n}^T=\lbrace 0,-1\rbrace.
\end{equation} We found that ${\bf R}=\mathds{1}$.

 \item  Solution corresponding to  $\kappa=1$  is given by 
 \begin{equation}
\mathbf{a}^T=\lbrace 0.436436, 2.26779\rbrace \quad \mathbf{n}^T=\lbrace 0.981981, -0.188982 \rbrace,
\end{equation} 
and  the corresponding rotation angle is 38.2132 degrees.. For $\kappa=-1$, the solution is 
\begin{equation}
\mathbf{a}^T=\lbrace -1.1547, 2.\rbrace \quad \mathbf{n}^T=\lbrace -0.866025, -0.5\rbrace.
\end{equation} 
 The corresponding rotation angle is $60$ degrees.

\item Solution corresponding to  $\kappa=1$  is given by 
  \begin{equation}
\mathbf{a}^T=\lbrace 1.1547005, 2\rbrace \quad \mathbf{n}^T=\lbrace 0.8660256, -0.5\rbrace,
\end{equation} 
 The corresponding rotation angle is $\pm120$ degrees. The solution corresponding to  $\kappa=-1$  is  
  \begin{equation}
\mathbf{a}^T=\lbrace -0.436436, 2.26779\rbrace \quad \mathbf{n}^T=\lbrace -0.9819805, -0.188982\rbrace,
\end{equation}  The corresponding rotation angle is $21.7868$ degrees.

\item Solution corresponding to  $\kappa=1$  is given by 
  \begin{equation}
\mathbf{a}^T=\lbrace 1.1547005, 2\rbrace \quad \mathbf{n}^T=\lbrace 0.866025, -0.5\rbrace,
\end{equation} 
 The corresponding rotation angle is $60$ degrees. The solution corresponding to  $\kappa=-1$  is  
  \begin{equation}
\mathbf{a}^T=\lbrace-0.436436, 2.26779,\rbrace \quad \mathbf{n}^T=\lbrace -0.9819805, -0.188982\rbrace,
\end{equation} the corresponding rotation angle is $38.2132$ degrees. 
\end{mycases}

The results given above suggest that micro-twinning is possible in triangular lattices since there are several cases for which the rotation $\mathbf{R}$ belongs to the point group of the triangular lattice. However, as opposed to the case of  square lattice, we did not observe any micro-twinning patterns in our numerical experiments in triangular lattices. One possible explanation   is the strong misalignment, in the case of triangular lattices between the orientation of the \textit{macro-modulations} and the lattice vectors when 	the critical loading is approached. Instead, in the case of square lattices we observe lattice scale modulations corresponding to the wave vectors at the boundary of the Brillouin zone present already in the original unstable mode, which is a perfect arrangement to generate a micro-laminate, see \cite{Salman2021-sn} for a detailed explanation on developing instability modes. 
	 
Second, we  study  the compatibility of the two nearest wells with the original triangular lattice that we take as identity ${\bf G}=\mathds{1}$. We have again 4 cases to consider (i) ${\bf H}=  \begin{pmatrix}
1 & \gamma^2\\
0 & 1 
\end{pmatrix}$,
(ii)  
${\bf H}=  \begin{pmatrix}
1 & -\gamma^2\\
0 & 1 
\end{pmatrix}$,
(iii)  
${\bf H}=  \begin{pmatrix}
0.5 & \sqrt{3}/6\\
-\sqrt{3}/2 & 1.5 
\end{pmatrix}$,
(iv)  
${\bf H}=  \begin{pmatrix}
0.5 &- \sqrt{3}/6\\
\sqrt{3}/2 & 1.5 
\end{pmatrix}$.
\begin{mycases}

\item   Solution corresponding to  $\kappa=1$  is given by 

\begin{equation}
\mathbf{a}^T=\lbrace 0.57735, 1. \rbrace \quad \mathbf{n}^T=\lbrace -0.866025, 0.5 \rbrace,
\end{equation} 
and the corresponding rotation angle is 60 degrees. . For $\kappa=-1$, the solution is given by
  \begin{equation}
\mathbf{a}^T=\lbrace 1.1547,0 \rbrace \quad \mathbf{n}^T=\lbrace 0., -1\rbrace.
\end{equation} We found that ${\bf R}=\mathds{1}$.

\item   Solution corresponding to  $\kappa=1$  is given by 
\begin{equation}
\mathbf{a}^T=\lbrace -0.57735, 1. \rbrace \quad \mathbf{n}^T=\lbrace 0.866025, -0.5 \rbrace,
\end{equation} 
and the corresponding rotation angle is 60 degrees. . For $\kappa=-1$, the solution is given by
  \begin{equation}
\mathbf{a}^T=\lbrace -1.1547,0 \rbrace \quad \mathbf{n}^T=\lbrace 0., -1\rbrace.
\end{equation} We obtain  ${\bf R}=\mathds{1}$.

\item Solution corresponding to  $\kappa=1$  is given by 
  \begin{equation}
\mathbf{a}^T=\lbrace 0.57735, 1\rbrace \quad \mathbf{n}^T=\lbrace 0.8660256, -0.5\rbrace,
\end{equation} 
We obtain  ${\bf R}=\mathds{1}$. The solution corresponding to  $\kappa=-1$  is  
  \begin{equation}
\mathbf{a}^T=\lbrace -0.57735, 1\rbrace \quad \mathbf{n}^T=\lbrace -0.8660256, -0.5\rbrace,
\end{equation}  The corresponding rotation angle is $60$ degrees.

\item Solution corresponding to  $\kappa=1$  is given by 
  \begin{equation}
\mathbf{a}^T=\lbrace 0.57735, 1\rbrace \quad \mathbf{n}^T=\lbrace 0.866025, -0.5\rbrace,
\end{equation} 
 The corresponding rotation angle is $60$ degrees. The solution corresponding to  $\kappa=-1$  is  
  \begin{equation}
\mathbf{a}^T=\lbrace-0.57735, 1\rbrace \quad \mathbf{n}^T=\lbrace -0.866025, -0.5\rbrace,
\end{equation} the corresponding rotation is  ${\bf R}=\mathds{1}$. 
\end{mycases}

\bigskip 
\bigskip

\end{appendices}

\section*{Bibliography}
\bibliographystyle{elsarticle-num}


\end{document}